%% file: main.tex
\newcommand{\DoPrePrint}{0} % 0 for 2-column submission/review format; 1 for double-spaced, line-numbered preprint
\newcommand{\sizecheck}{0} %removes title, abstract, author list, acknowledgements and references
\newcommand{\DoColor}{1} %set to 1 to get color figures for arxiv.

\ifnum\DoPrePrint=1
  \RequirePackage{lineno}  % LINENO  CTOBS
  \documentclass[aps,preprint,showpacs,superscriptaddress]{revtex4-1} % CTOBS
  \usepackage{lineno}
\else
  \documentclass[aps,prl,twocolumn,showpacs,superscriptaddress,nofootinbib]{revtex4-1}  % UTOBS
\fi

\usepackage{graphicx}  % needed for figures
\usepackage{dcolumn}   % needed for some tables
\usepackage{bm}        % for math
\usepackage{amssymb}   % for math
\usepackage{hyperref}
\usepackage{xspace}

\newcommand{\minerva}{MINERvA\xspace}
\newcommand{\Xbj}{\ensuremath{x}\xspace}
\newcommand{\Gev}{GeV\xspace}
\newcommand{\GevSq}{GeV$^{2}$\xspace}

\ifnum\DoColor=1
\graphicspath{{./figures/}}
\else
\graphicspath{{./figures_gs/}}
\fi

\begin{document}

% the following line is for submission, including submission to the arXiv!!
%\hspace{5.2in} \mbox{Fermilab-Pub-15-556-ND}

\title{Measurement of Partonic Nuclear Effects in Deep-Inelastic Neutrino Scattering using \minerva}
\input{authorlist.tex}       
\date{\today}

\ifnum\sizecheck=0
\begin{abstract}
The \minerva collaboration reports a novel study of neutrino-nucleus charged-current deep inelastic scattering (DIS) using the same neutrino beam incident on targets of polystyrene, graphite, iron, and lead. Results are presented as ratios of C, Fe, and Pb to CH. The ratios of total DIS cross sections as a function of neutrino energy and flux-integrated differential cross sections as a function of the Bjorken scaling variable \Xbj are presented in the neutrino-energy range of $5-50$ GeV. Based on the predictions of charged-lepton scattering ratios, good agreement is found between the data and prediction at medium x and low neutrino energy. However, the ratios appear to be below predictions in the vicinity of the nuclear shadowing region, $\Xbj < 0.1$. This apparent deficit, reflected in the DIS cross-section ratio at high $E_{\nu}$, is consistent with previous \minerva observations [B. Tice {\it et al.} (MINERvA Collaboration), Phys. Rev. Lett. {\bf 112}, 231801 (2014)] and with the predicted onset of nuclear shadowing with the the axial-vector current in neutrino scattering. 
\end{abstract}
\fi


\maketitle

\ifnum\DoPrePrint=1
\linenumbers
\fi

Deep inelastic scattering (DIS) has played an important role in the history of physics \cite{heraOverview}. Starting with the confirmation of the quark parton model \cite{slacQuarks}, high-energy DIS experiments, mainly using charged leptons (muons and electrons) as probes, have been essential tools in understanding parton dynamics~\cite{Geesaman:1995yd}. These experiments have also contributed to the development of perturbative Quantum Chromodynamics (QCD) that describes the quark and gluon content of the nucleon over a broad kinematic regime.

%Deep inelastic scattering (DIS) has played an important role in the history of physics \cite{heraOverview}. Starting with the confirmation of the quark parton model \cite{slacQuarks}, high-momentum transfer and high-energy experiments have been essential tools in understanding parton dynamics~\cite{Geesaman:1995yd}. The development of the quark parton model were important to the gradual discovery of quantum chromodynamics (QCD) to explain observed scaling violations \cite{scalingViolation}. Precision QCD measurements of quarks and gluons have also been made possible using electron-proton DIS \cite{Collaboration:2010ry}. Traditionally, these experiments have used charged leptons (muons and electrons) as probes.

Charged-lepton DIS has been used as a parton-level tool for exploring nuclear effects on a variety of targets \cite{leptonNukeEff}. These effects are typically parameterized as a function of four-momentum transfer squared $Q^{2} = -q^{2}$ and the Bjorken scaling variable $\Xbj$ \cite{bjorkenScaling}, the fraction of the nucleon's momentum carried by the struck parton in the infinite momentum frame: 

\begin{equation}
\label{eqn:xBj}
\Xbj = \frac{Q^{2}}{2M_{N}E_{had}},
\end{equation}

\noindent where $M_{N}$ is the average nucleon mass, $M_{N} = (M_{p}+M_{n})/2$, and $E_{had}$ is the total energy of the final state hadrons. In charged-lepton experiments $E_{had}$ is typically replaced by $\nu$, the energy loss of the incident lepton. As this energy loss cannot be measured directly in a neutrino beam, $E_{had}$ is used as an estimator. Four distinct effects have been identified in the ratios of total DIS and differential DIS cross sections per nucleon on heavy nuclei such as iron, gold and calcium \cite{slacRatios}, to those on deuterium. At $\Xbj \lesssim 0.1$, ``shadowing'' depletes the bound cross section \cite{shadowing}, while anti-shadowing produces a compensating increase for $0.1 \lesssim \Xbj \lesssim 0.3$ \cite{Aubert:1981gw}. The EMC effect in the region $0.3 \lesssim \Xbj \lesssim 0.75$ reduces the bound cross section \cite{emcOne, emcTwo}, and Fermi motion, dominant at $\Xbj \gtrsim 0.75$,  causes a sharp enhancement of the bound cross section \cite{Bodek:1980ar}. Nuclear shadowing and Fermi motion are fairly well understood theoretically and experimentally. Anti-shadowing is assumed to compensate the dips in the shadowing and EMC region. However, the EMC effect currently has no widely-accepted theoretical origin \cite{cloetEMC}. 

Nuclear effects in neutrino-induced DIS are much less explored. To date no partonic nuclear effects, similar to those measured for charged-lepton DIS, have been directly measured due to the difficulty in combining data sets with different neutrino fluxes, acceptances, thresholds, and resolutions. The analyses that do exist measure neutrino DIS in heavy nuclei such as Fe \cite{Berge:1989hr, ccfr, Fleming:2000bg, PhysRevD.74.012008}, Ne \cite{Varvell:1987qu}, and Pb \cite{onegut200665}. Comparing measurements in heavy nuclei to free-nucleon calculations in an attempt to determine neutrino-nuclear effects has shown some tension with charged-lepton nuclear effects \cite{Kovarik:2010uv}. Due to these unresolved inconsistencies, the typical approach for modern neutrino DIS models has been to adapt existing charged-lepton nuclear effects into neutrino DIS models \cite{BY}. 

This paper presents a measurement of nuclear effects in charged-current neutrino DIS using the \minerva detector. A previous analysis with the \minerva detector and nuclear targets of inclusive ratios contained a large percentage of resonant (approximately $35 \%$) and quasi-elastic ($11 - 50 \%$) events \cite{nukePRL}, that do not allow the data to be interpreted at the parton level. While neutrino experiments present many challenges, including knowledge of the neutrino flux and the unknown effect of final-state interactions, neutrinos provide a unique weak-only probe of the atomic nucleus. There is no \emph{a priori} reason to assume neutrino and charged-lepton DIS will be identical, as neutrinos are uniquely sensitive to both the axial vector and vector components of the weak nuclear force \cite{Kulagin2006126}.    

The \minerva experiment, as well as many other current \cite{novaProp, argoneut, microBoone} neutrino experiments, uses the \textsc{genie} event generator \cite{genie} to simulate neutrino interactions in the detector. This generator is used to simulate the signal DIS as well as the background quasielastic interactions, resonance production and the transition region from resonant to DIS events. \textsc{genie}'s simulation of DIS and transition events is based on the 2003 Bodek-Yang model \cite{BY}, which computes cross sections at the partonic $\nu_{\mu} +$~quark level using GRV98LO PDFs \cite{grvnlo} to calculate the structure functions $F_{2}$, and $\Xbj F_{3}$. The structure function $2\Xbj F_{1}$ is related to $F_{2}$ via the ratio of the transverse ($\sigma_{T}$) to longitudinal ($\sigma_{L}$) cross-sections $R_{L} = \sigma_{L} / \sigma_{T}$: 

\begin{equation}
2\Xbj F_{1} = \frac{1+Q^{2}/E^{2}_{had}}{1+R_{L}}F_{2}.
%2xF_1 = \frac{1 + Q^2/E_had^2}{1+R_L} F_2
\end{equation}  

\noindent \textsc{genie} uses the Whitlow parameterization \cite{whitlow} for $R_{L}$. Bodek-Yang accounts for target-mass modification and higher-twist effects by calculating the nucleon structure functions as a function of a modified scaling variable \cite{BY}. Coefficients of this scaling variable are tuned to data from a variety of charged-lepton scattering experiments, and the uncertainties on these fits are propagated to the analysis. The nuclear modification made to the structure functions is applied identically to all elements heavier than helium. \textsc{genie}'s predicted total DIS and differential cross sections of carbon, polystyrene scintillator (CH), iron, and lead are identical once the differing neutron fractions are taken into account. This treatment does not take account of the $A$-dependence of shadowing and the EMC effect established in charged-lepton scattering \cite{Kopeliovich:1995qg, Arrington:2012ax}.

The \minerva neutrino detector is deployed in the NuMI \cite{Adamson:2015dkw} neutrino beam at the Fermi National Accelerator Laboratory, approximately 1 km away from the neutrino production target. The broad-band neutrino energy spectrum peaks at approximately 3~GeV, however it extends to above 100~GeV. The generation of mesons produced from p + C collisions inside a graphite target is modeled using Geant4 \cite{geant}. External data from NA49 \cite{na49} and MIPP \cite{mippPaper} are used to constrain and improve the simulation for neutrino energies below $30$ GeV, via reweighting the default Geant4 prediction \cite{Fields:2013zhk} \footnote{This paper uses the ``Generation 1" \minerva flux to calculate background rates.}.

The \minerva detector, detailed in Ref \cite{detNim}, uses hexagonal planes made up of triangular scintillator strips for charged-particle tracking and for reconstruction of hadronic and electromagnetic showers. The most upstream region contains passive nuclear targets of solid graphite, iron, and lead, each with upstream and downstream scintillator planes to provide tracking, vertexing and shower reconstruction between the targets. Downstream of the nuclear targets are fully-active tracker scintillator planes and electromagnetic and hadronic calorimeters. Each of these regions is surrounded by an outer electromagnetic calorimeter as well as an outer detector used for side-exiting hadronic calorimetry. The magnetized MINOS detector \cite{minosNim}, located 2~m downstream of \minerva, serves as a muon spectrometer.

Charged-current $\nu_{\mu}$ DIS is characterized by a final state consisting of a $\mu^{-}$ and a hadronic shower with invariant mass above the resonance region. All deposits of energy in the \minerva detector are sorted into spatially associated ``clusters" within each plane. Collinear clusters are used to reconstruct particle trajectories (tracks) through the passive nuclear targets, tracker, and calorimeter regions. Tracks in MINOS are identified in a manner described in \cite{MinosTracking}. The longest track in the recorded interaction matched to a track in MINOS is identified as the primary muon, and all other clusters in \minerva are identified as the hadronic shower. MINOS matching limits the angular acceptance of events, and only muons that are within 17$^{\circ}$ of the beam direction are included. The charge sign and momentum of the muons are measured by the MINOS near detector.    

After reconstructing all available tracks, an event is assigned a vertex using an iterative Kalman \cite{kalman} fitter when multiple tracks are present. Approximately $20 \%$ of DIS events contain only one identified track with additional untracked energy, in which case the vertex is reconstructed to the start point of the track. In order to capture single-track events originating from the nuclear targets, the event selection allows vertices originating in two scintillator planes downstream and one plane upstream to be included in the target sample in both single and multi-track events. This leads to a background of non-nuclear target events which is subtracted as described below.

The DIS sample is isolated using kinematic selections based on the $Q^{2}$ and invariant mass $W$ of the recoil system. Both quantities are calculated from the muon energy $E_{\mu}$, the outgoing muon angle $\theta_{\mu}$, and $E_{had}$:

\begin{eqnarray}
\label{eqn:nukeEff}
Q^{2} = 4E_{\nu} E_{\mu} \sin^{2} \left( \frac{\theta_{\mu}}{2} \right ),\\ \nonumber
W^{2} = M_{N}^{2}+2M_{N}E_{had} - Q^{2},
\end{eqnarray}

\noindent where the reconstructed neutrino energy is equal to the sum of the muon and hadronic energy, $E_{\nu} = E_{\mu} + E_{had}$.  DIS signal events are required to have $Q^{2} \geq 1.0$ \GevSq and $W \geq 2.0$ \Gev. The $Q^{2}$ of these events is sufficiently large to resolve the nucleon into its parton constituents. The selection of high-$W$ events serves to remove quasielastic and resonant interactions from the sample.  

The selected event sample contains two backgrounds, both of which are subtracted bin-by-bin from candidate event distributions. The first type arises from detector effects, smearing low $W$ and $Q^{2}$ events upward into the DIS selection. The rate of these events is estimated by scaling the Monte Carlo (MC) simulation to agree with data in two sidebands: $Q^{2} \geq 1.0$ \GevSq, $1.3 \leq W < 1.8$ \Gev and $Q^{2} < 0.8$ \GevSq, $W \geq 2.0$ \Gev. The data in these regions are used to tune two background templates. The first template contains all simulated events with \emph{generated} $W_{gen} < 2.0$ \Gev (``low $W$"), and the second consists of events with a generated $W_{gen} > 2.0$ \Gev and $Q_{gen}^{2} < 1.0$ \GevSq (``low $Q^{2}$"). The low-$W$ template includes the quasielastic and resonant events. The normalization of each template is fit to the data simultaneously in both sidebands for each nucleus over the energy range $5.0 \leq E_{\nu} < 50$ GeV. The fit results are summarized in Table \ref{tab:sbWeight}. The data tend to prefer a higher background rate at low $Q^{2}$.

\squeezetable
\begin{table}[h]
\begin{ruledtabular}
\begin{tabular}{ccc}
Target Material & Low $W$ & Low $Q^{2}$\\
\hline
CH & $0.94 \pm 0.01$  & $1.57 \pm 0.02$\\
C & $0.90 \pm 0.08$   & $1.58 \pm 0.11$\\
Fe & $0.99 \pm 0.04$  & $1.58 \pm 0.05$\\
Pb & $0.95 \pm 0.03 $ & $1.36 \pm 0.05$\\
\end{tabular}
\end{ruledtabular}
\caption{\label{tab:sbWeight} Scale factors applied to the two background templates. Low $W$: $W_{gen} < 2.0$ \Gev. Low $Q^{2}$: $W_{gen} > 2.0$ \Gev and $Q_{gen}^{2} < 1.0$ \GevSq. The uncertainties are the statistical uncertainties on the fit. Systematic uncertainties on the fits are evaluated by adjusting the underlying theoretical parameters of the simulation by $\pm 1 \sigma$ and re-running the fits.}
\end{table} 

\begin{figure}[h!]
\includegraphics[scale=0.5]{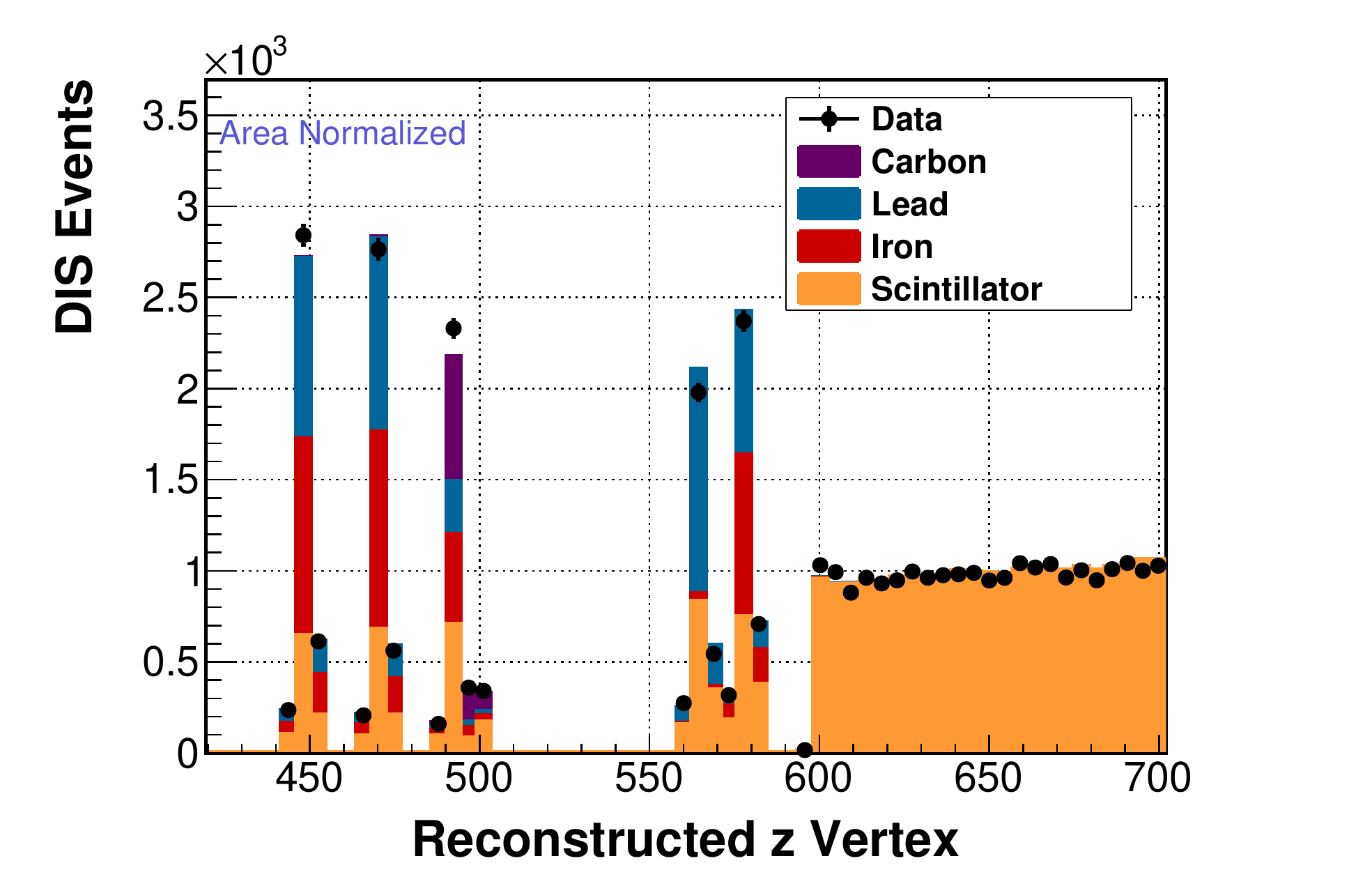}

\caption{\label{fig:zVertex} The number of DIS events in the passive nuclear targets ($0 < z < 600$ cm) and tracker region ($z > 600$ cm) as a function of longitudinal position. The yellow area in the first five peaks represents the scintillator background in each nuclear target. The events located in the scintillator between the passive targets are not shown in this Figure for clarity.}
\end{figure}

A second background arises from events mis-reconstructed in the passive nuclear target modules that originate in the scintillator modules surrounding the targets but are mis-reconstructed as originating in the passive nuclear target modules. Figure \ref{fig:zVertex} illustrates the simulation of the CH background as well as the passive target signal. These background events are subtracted by measuring the event rate of reconstructed DIS events in the \minerva tracker region in a manner similar to that described in \cite{nukePRL}. The nuclear target region is farther away from MINOS than the fully-active region and as a result the muon acceptances are different. A Geant4 simulation is used to evaluate the different acceptances. This procedure does not fully reproduce the simulated CH background and the difference between the estimated and true CH background in the simulation is included as an additional systematic uncertainty.   

\begin{figure}[h]
\includegraphics[scale=0.5]{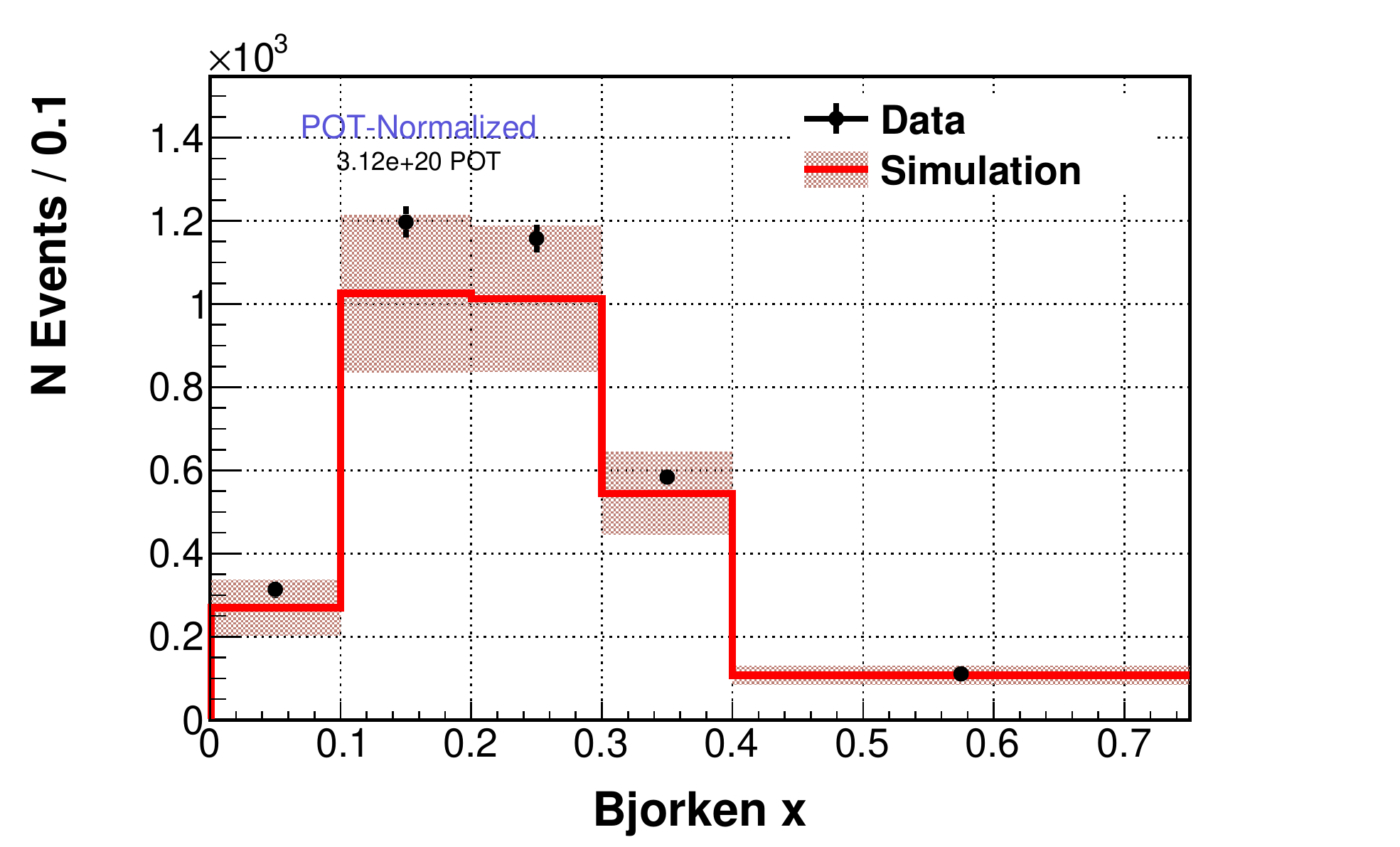}
\caption{\label{fig:xFe} Deep inelastic scattering events in iron as a function of unsmeared Bjorken-x. The total systematic uncertainty is drawn as a band around the simulation, and the error bars around the data are statistical.}
\end{figure}

Figure~\ref{fig:xFe} shows the distribution of DIS events in data and simulation in iron after applying all background corrections and unfolding to correct for detector smearing. The unfolding is based on Bayesian unfolding \cite{unfolding} with one iteration, which reduces biases in the unfolded distributions to the few-percent level. The migration matrix used in unfolding defines the unsmeared \Xbj based on the generated energy loss of the neutrino $\nu$, and corrects for the differences based on the  $E_{had}$ calculation. Systematic uncertainties at the level of $20 \%$ exist primarily due to the neutrino flux estimate. To mitigate flux uncertainties, and to directly evaluate partonic nuclear effects, ratios of cross sections are taken between the nuclear targets (C, Fe, Pb) and CH. Before taking ratios, detector efficiency and loss of DIS events due to $W$ and $Q^{2}$ smearing are corrected via an acceptance correction derived from the simulation. The acceptance correction does not include muons with angles greater than $17^{\circ}$ in either the total or differential cross section. This corresponds to a region of phase space where acceptance into the MINOS detector is poor, and the efficiency is low.

\begin{figure}
\centering
\ifnum\DoPrePrint=0
\includegraphics[trim = 0mm 10mm 25mm 8mm, clip, width=0.48\columnwidth]{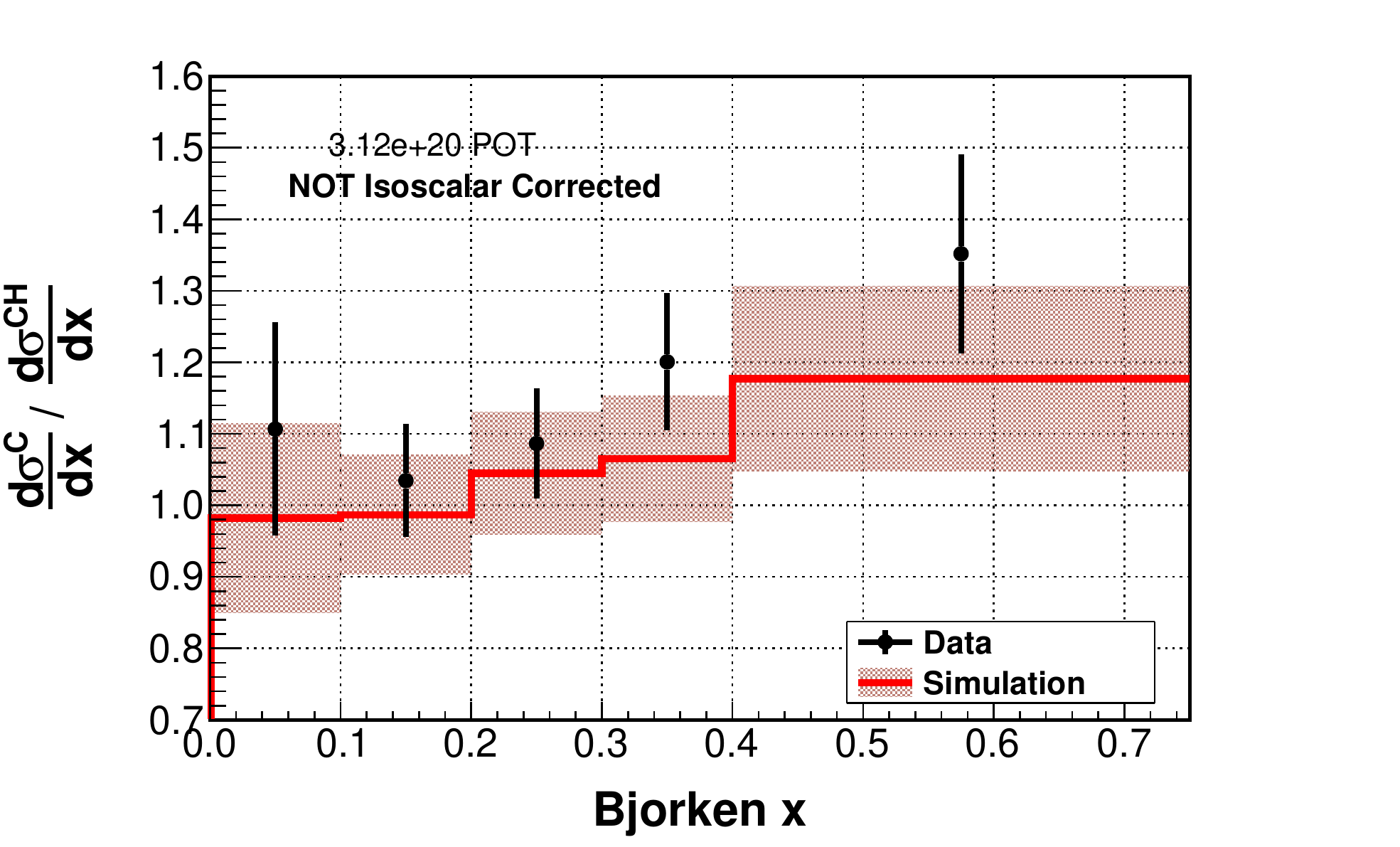}
\includegraphics[trim = 2mm 8mm 15mm 8mm, clip, height=2.45cm, width=0.48\columnwidth]{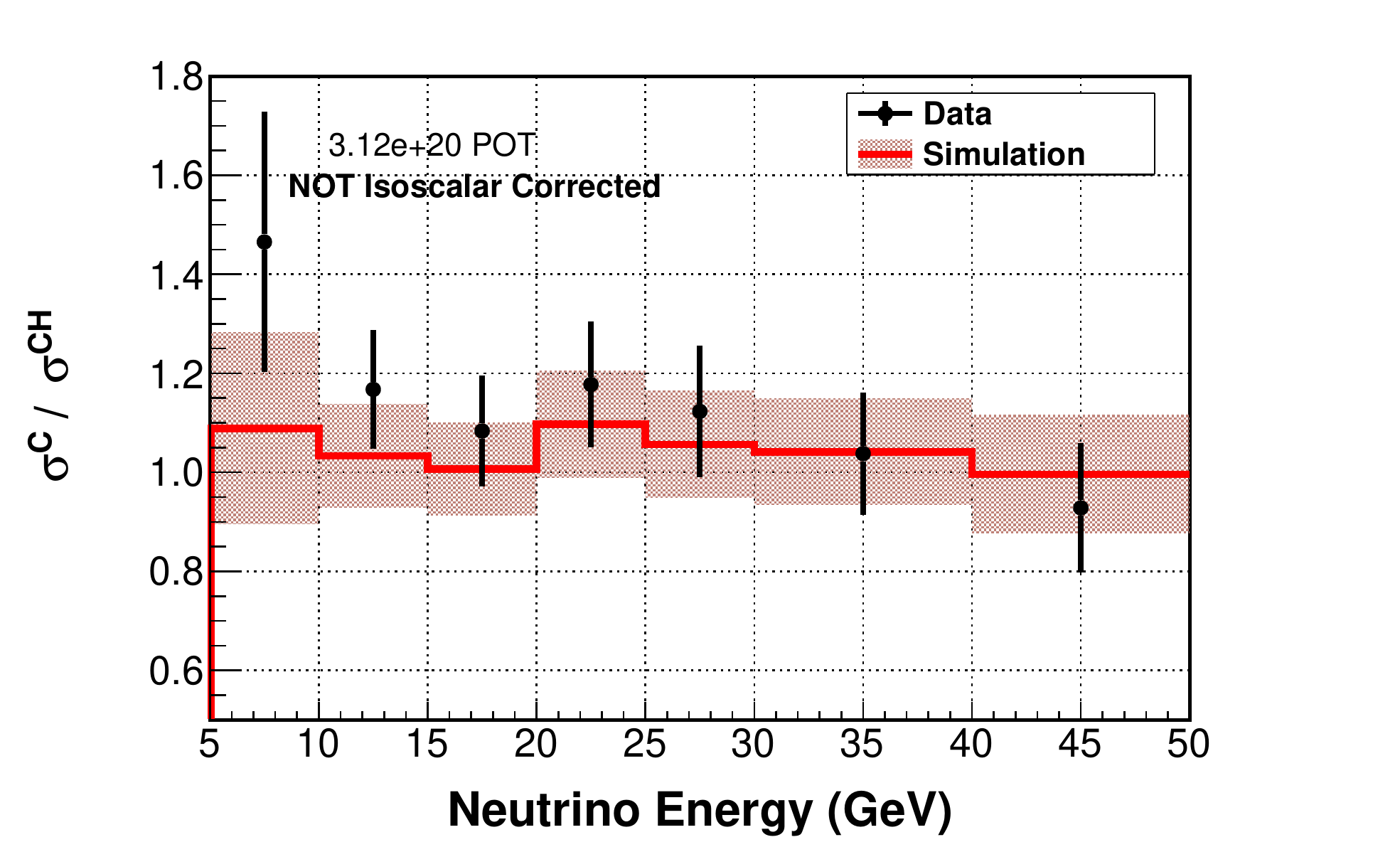}
\includegraphics[trim = 0mm 10mm 25mm 8mm, clip, width=0.48\columnwidth]{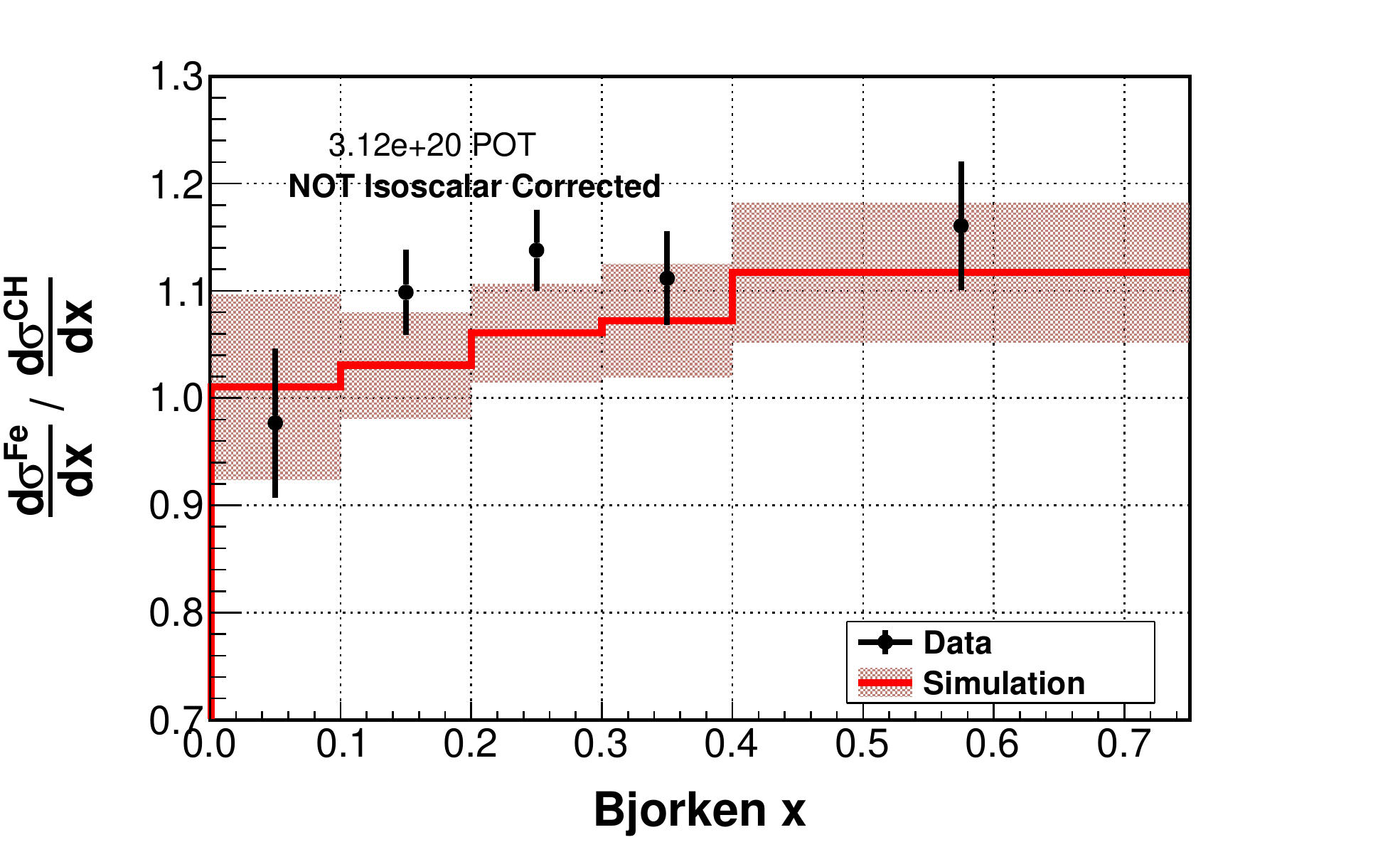}
\includegraphics[trim = 2mm 8mm 15mm 8mm, clip, height=2.45cm, width=0.48\columnwidth]{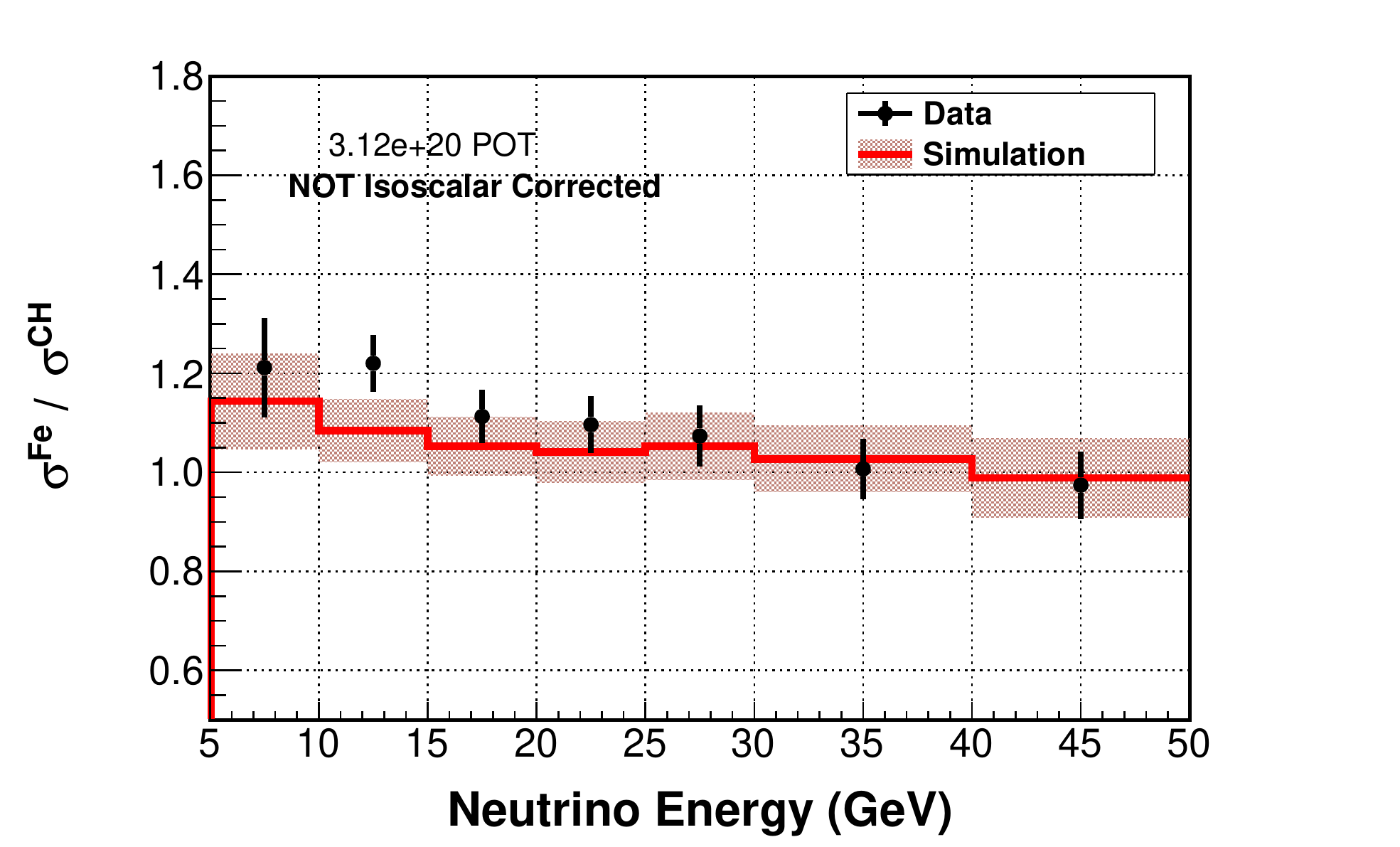}
\includegraphics[trim = 0mm 00mm 25mm 8mm, clip, width=0.48\columnwidth]{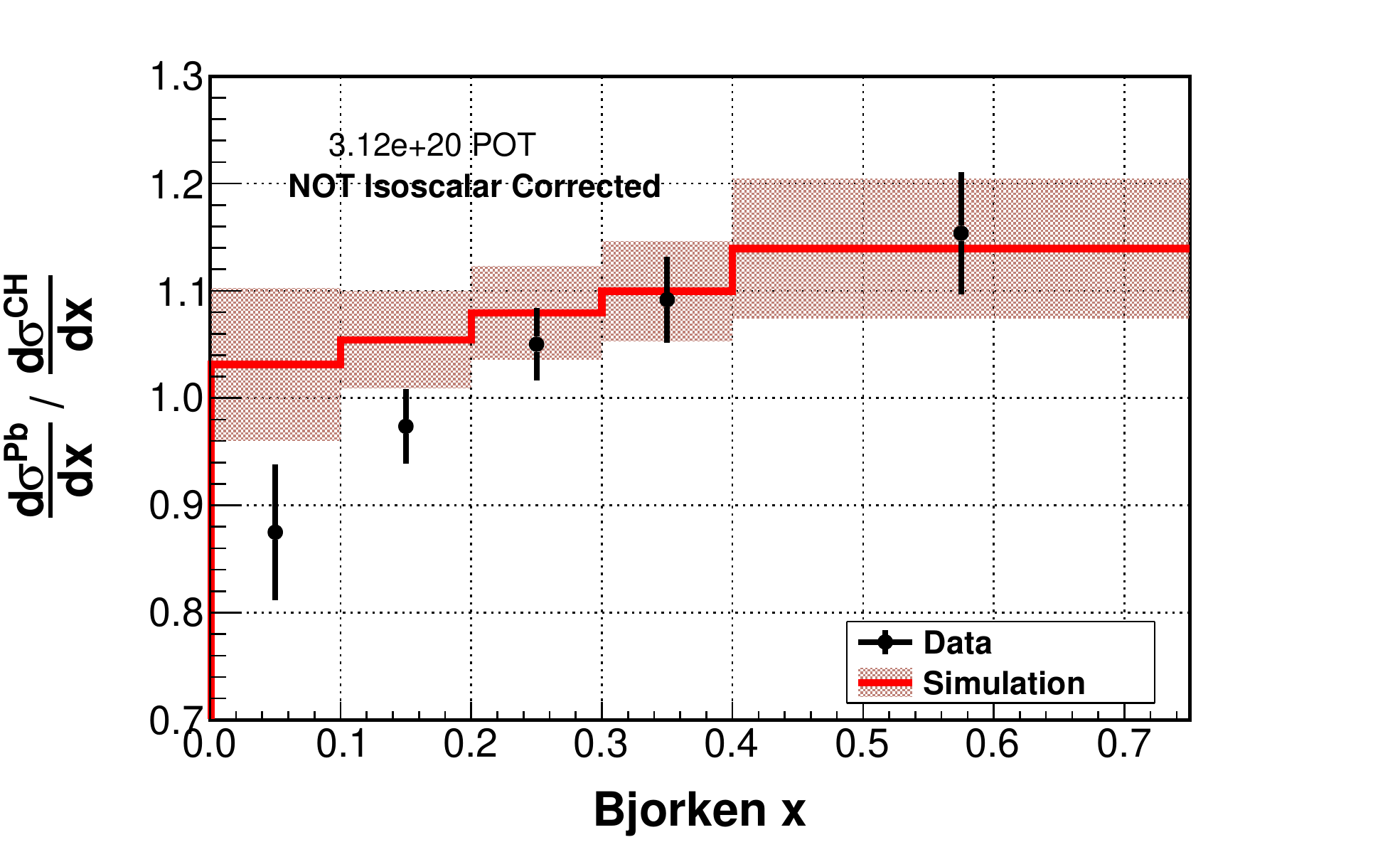}
\includegraphics[trim = 2mm 00mm 15mm 8mm, clip, height=2.65cm, width=0.48\columnwidth]{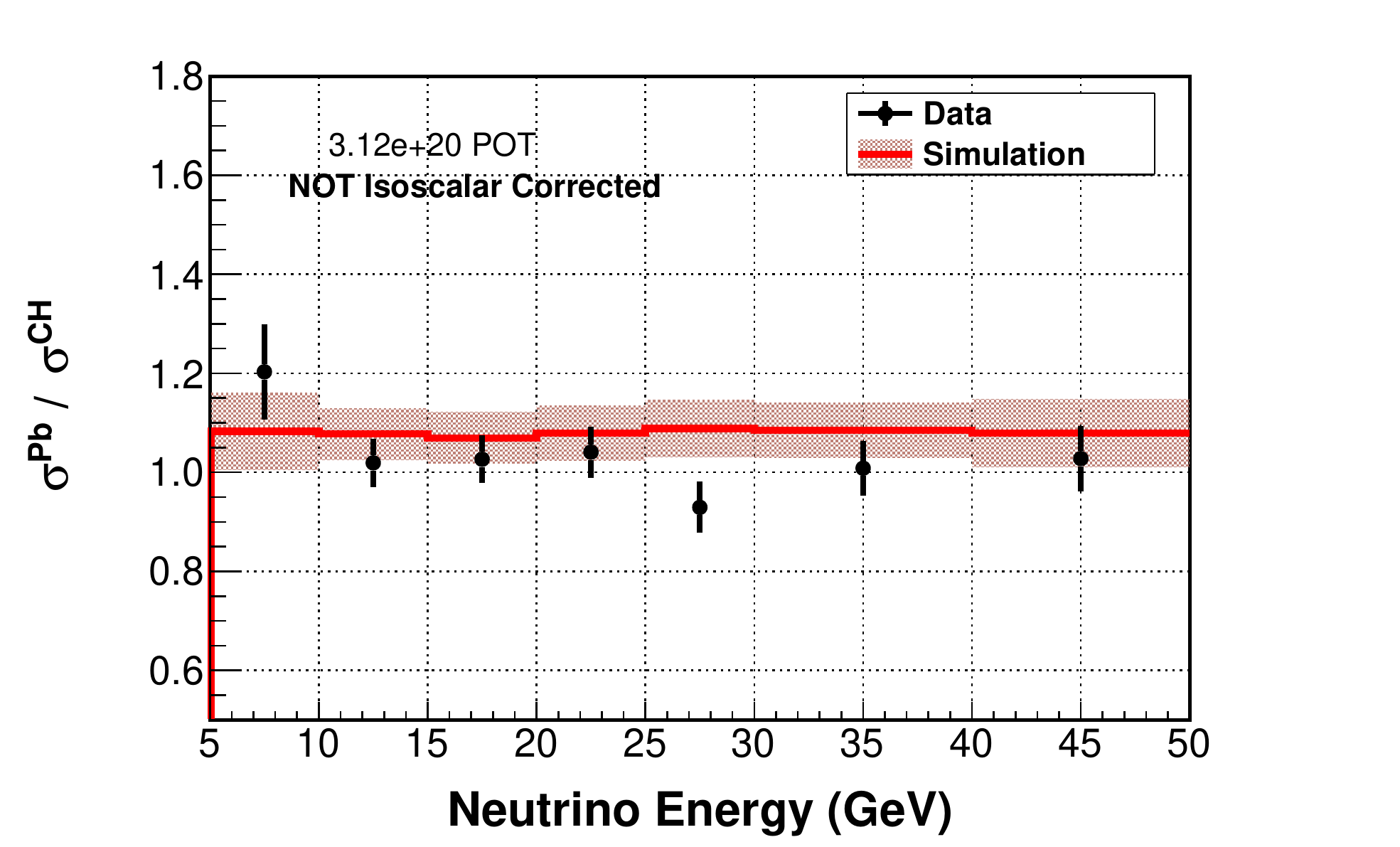}
\else
\includegraphics[trim = 0mm 10mm 25mm 8mm, clip, width=0.48\columnwidth]{RatioResultPlot_CrossSectionRatio_x_z06_91_Nu_v22_prl.pdf}
\includegraphics[trim = 2mm 8mm 15mm 8mm, clip, height=4.6cm,  width=0.48\columnwidth]{RatioResultPlot_CrossSectionRatio_Enu_z06_91_Nu_v22_prl.pdf}
\includegraphics[trim = 0mm 10mm 25mm 8mm, clip, width=0.48\columnwidth]{RatioResultPlot_CrossSectionRatio_x_z26_92_Nu_v22_prl.pdf}
\includegraphics[trim = 2mm 8mm 15mm 8mm, clip, height=4.6cm, width=0.48\columnwidth]{RatioResultPlot_CrossSectionRatio_Enu_z26_92_Nu_v22_prl.pdf}
\includegraphics[trim = 0mm 00mm 25mm 8mm, clip, width=0.48\columnwidth]{RatioResultPlot_CrossSectionRatio_x_z82_93_Nu_v22_prl.pdf}
\includegraphics[trim = 2mm 00mm 15mm 8mm, clip, height=5.1cm, width=0.48\columnwidth]{RatioResultPlot_CrossSectionRatio_Enu_z82_93_Nu_v22_prl.pdf}
\fi
\caption{Left: Ratio of the $\Xbj$-differential DIS cross section on C (top), Fe (center) and Pb (bottom) to CH.  Right: Ratio of the total DIS cross section on C (top), Fe (center) and Pb (bottom) to CH as a function of $E_{\nu}$. Data are drawn as points with statistical uncertainty and simulation as lines in both cases. The total systematic error is drawn as a band around the simulation in each histogram. \label{fig:crossSections} }
\end{figure}

\begin{figure}
\centering
\ifnum\DoPrePrint=0
\includegraphics[trim = 0mm 10mm 25mm 8mm, clip, width=0.48\columnwidth]{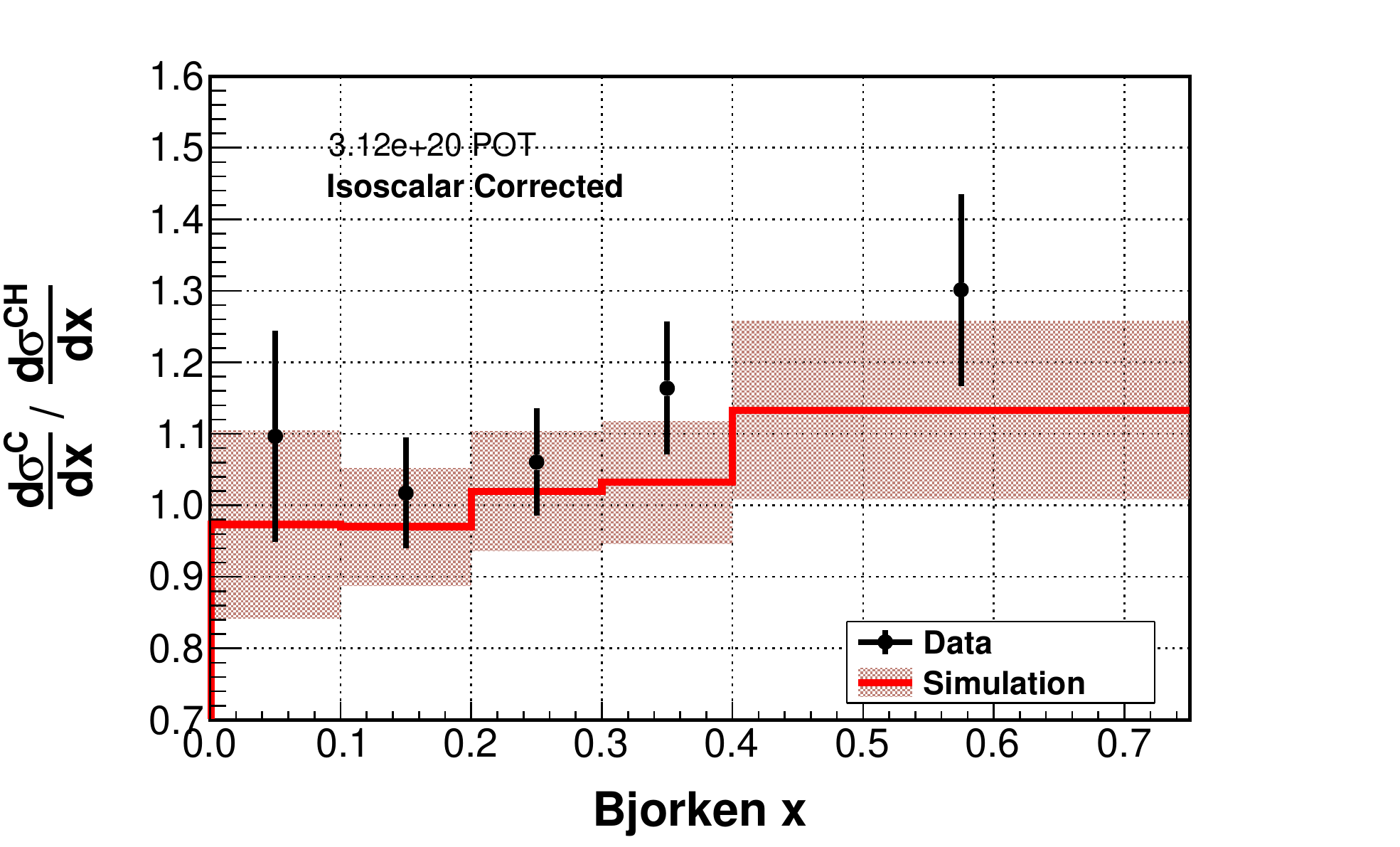}
\includegraphics[trim = 2mm 8mm 15mm 8mm, clip, height=2.45cm, width=0.48\columnwidth]{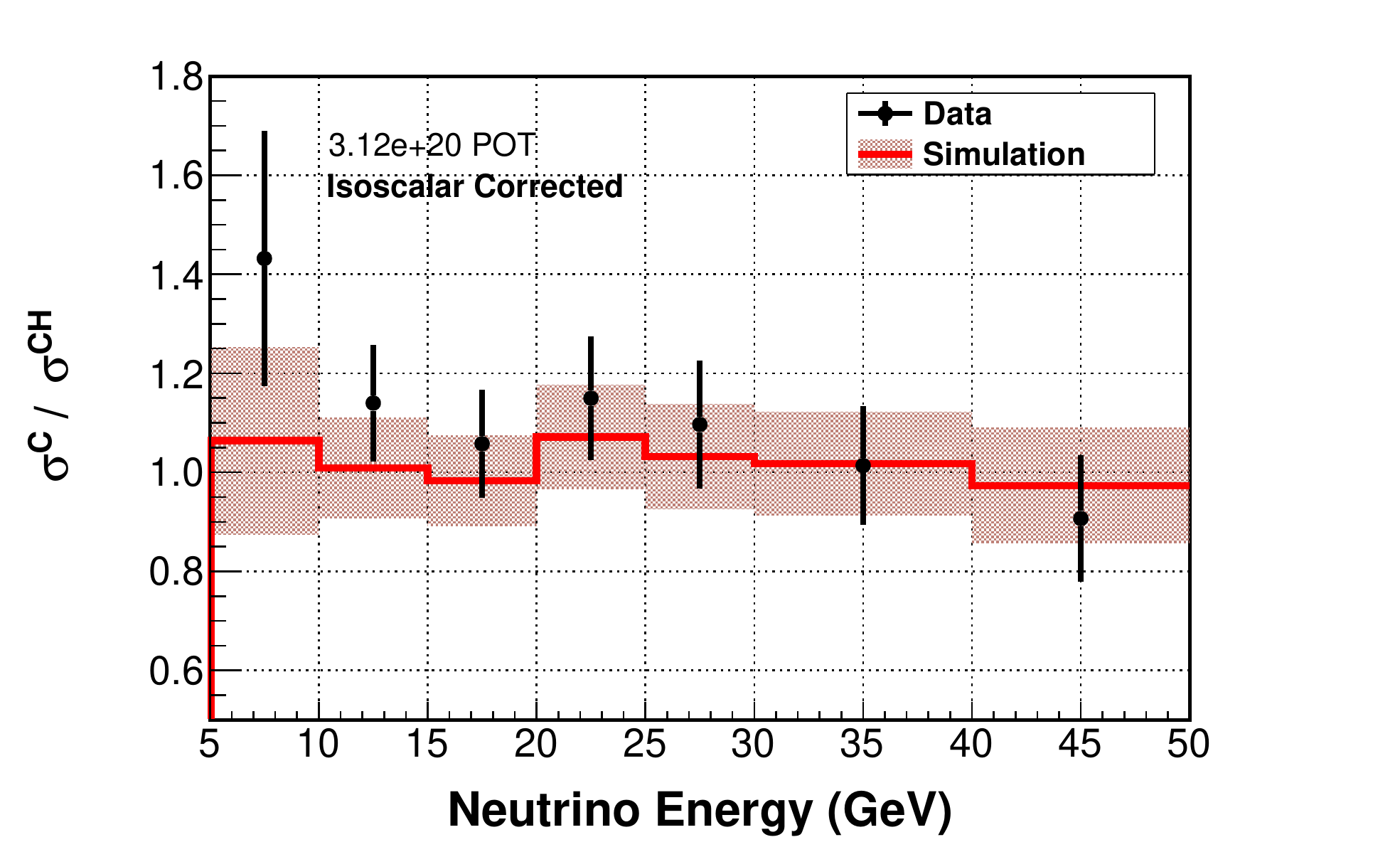}
\includegraphics[trim = 0mm 10mm 25mm 8mm, clip, width=0.48\columnwidth]{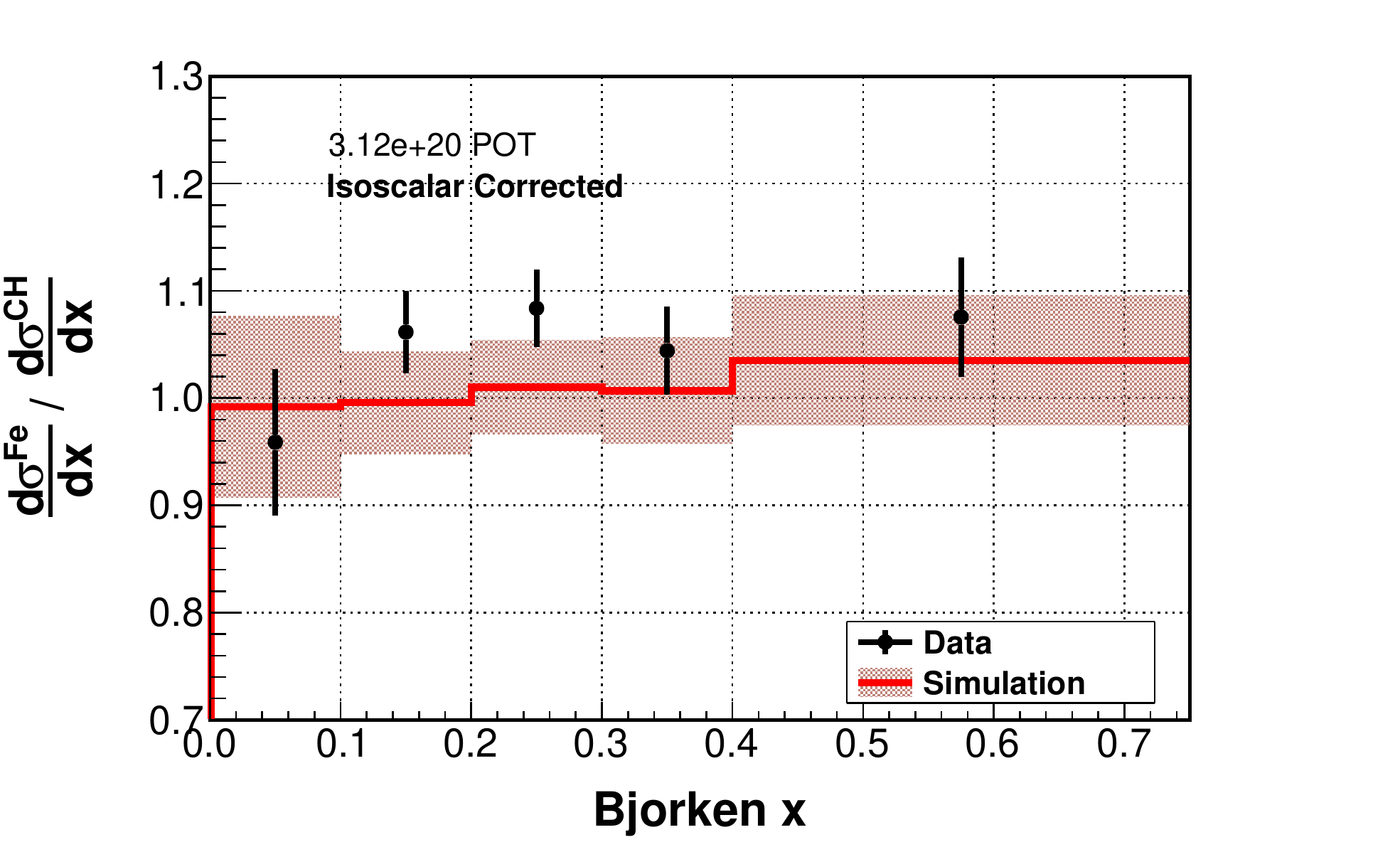}
\includegraphics[trim = 2mm 8mm 15mm 8mm, clip, height=2.45cm, width=0.48\columnwidth]{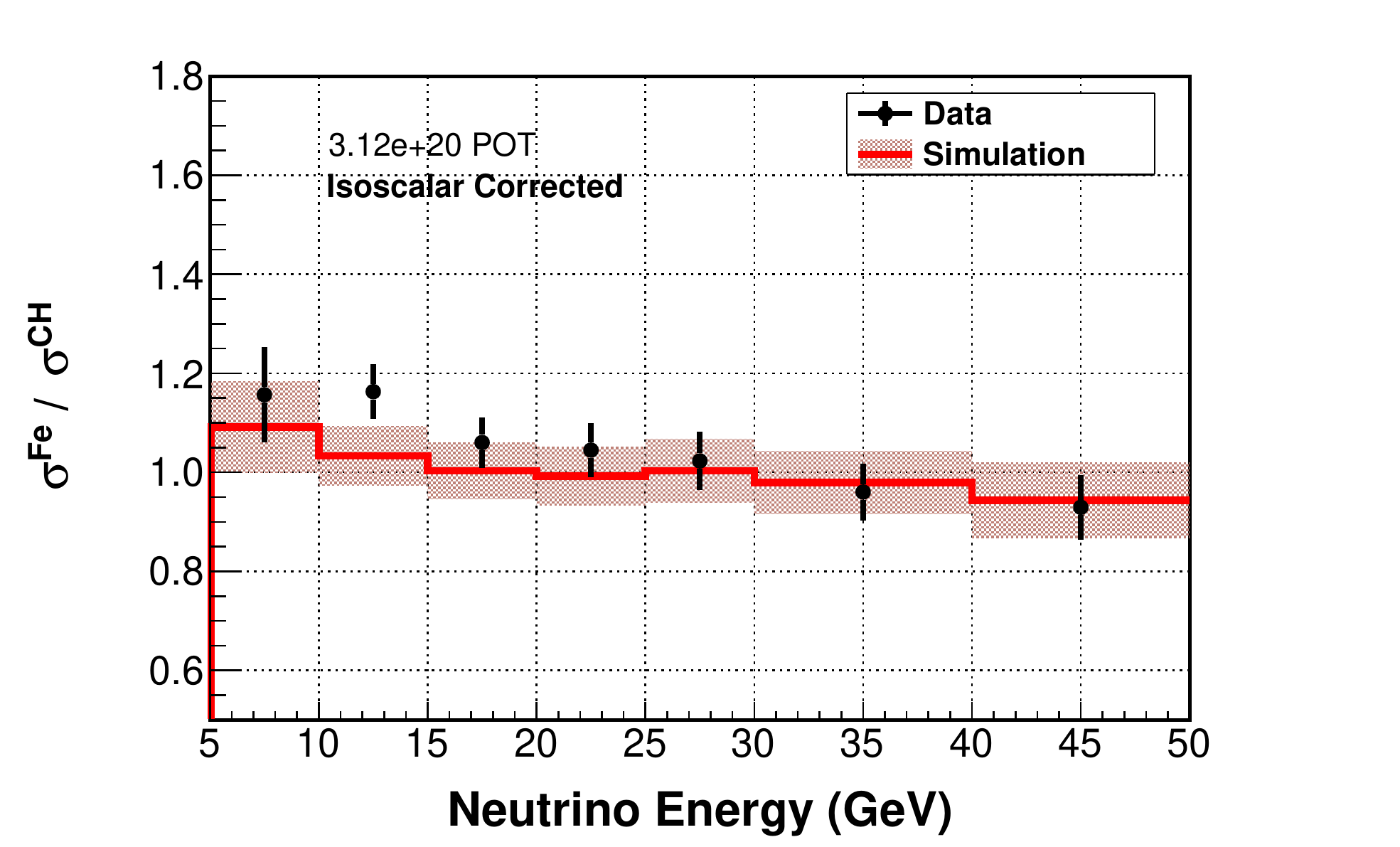}
\includegraphics[trim = 0mm 00mm 25mm 8mm, clip, width=0.48\columnwidth]{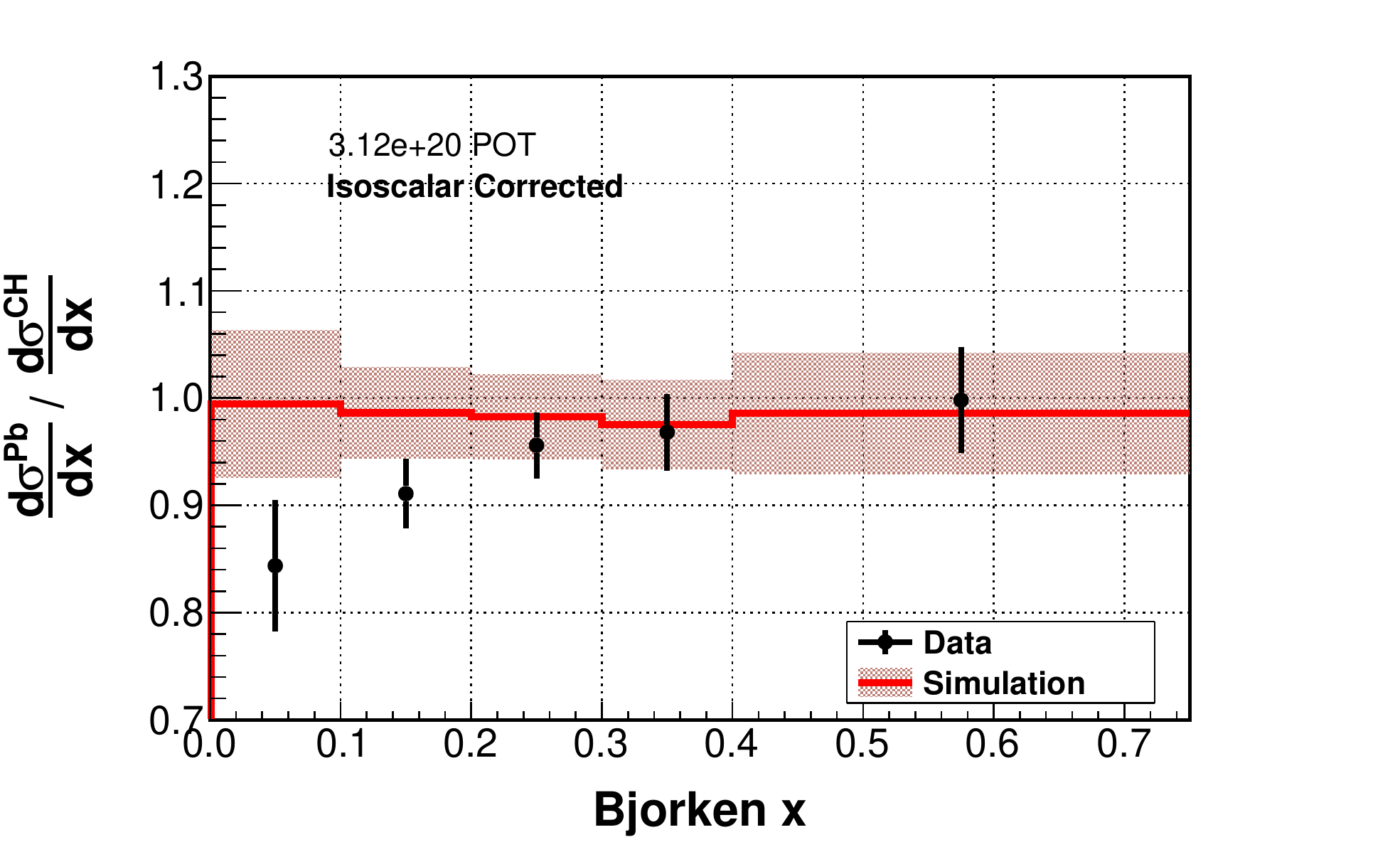}
\includegraphics[trim = 2mm 00mm 15mm 8mm, clip, height=2.65cm, width=0.48\columnwidth]{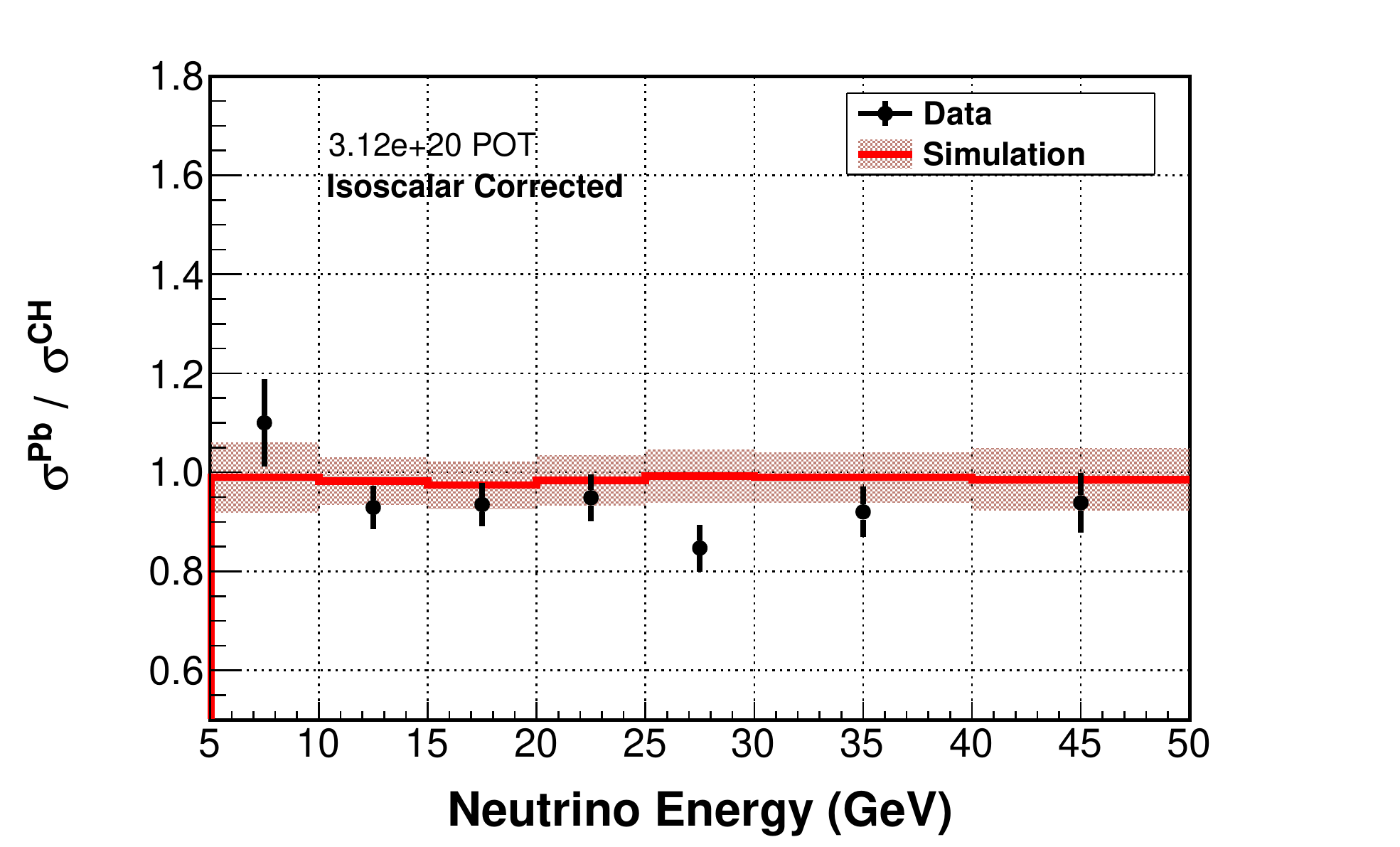}
\else
\includegraphics[trim = 0mm 10mm 25mm 8mm, clip, width=0.48\columnwidth]{RatioResultPlot_CrossSectionIsoRatio_x_z06_91_Nu_v22_prl.pdf}
\includegraphics[trim = 2mm 8mm 15mm 8mm, clip, height=4.6cm,  width=0.48\columnwidth]{RatioResultPlot_CrossSectionIsoRatio_Enu_z06_91_Nu_v22_prl.pdf}
\includegraphics[trim = 0mm 10mm 25mm 8mm, clip, width=0.48\columnwidth]{RatioResultPlot_CrossSectionIsoRatio_x_z26_92_Nu_v22_prl.pdf}
\includegraphics[trim = 2mm 8mm 15mm 8mm, clip, height=4.6cm, width=0.48\columnwidth]{RatioResultPlot_CrossSectionIsoRatio_Enu_z26_92_Nu_v22_prl.pdf}
\includegraphics[trim = 0mm 00mm 25mm 8mm, clip, width=0.48\columnwidth]{RatioResultPlot_CrossSectionIsoRatio_x_z82_93_Nu_v22_prl.pdf}
\includegraphics[trim = 2mm 00mm 15mm 8mm, clip, height=5.1cm, width=0.48\columnwidth]{RatioResultPlot_CrossSectionIsoRatio_Enu_z82_93_Nu_v22_prl.pdf}
\fi
\caption{Left: Isoscalar-corrected ratios of the $\Xbj$-differential DIS cross section on C (top), Fe (center) and Pb (bottom) to CH. Right: Ratio of the total DIS cross section on C (top), Fe (center) and Pb (bottom) to CH. Data are drawn as points with statistical uncertainty and simulation as lines in both cases. The total systematic error is drawn as a band around the simulation in each histogram. \label{fig:crossSectionsIso}}
\end{figure}

The differential cross section ratios after applying the background corrections are shown in Fig.~\ref{fig:crossSections} (left). A breakdown of uncertainties for the differential ratios may be found in Table \ref{tab:x_ratio_sys_errors}. There is an $\Xbj$-dependence to the ratios due to the neutron excesses in Fe and Pb. This manifests itself as an increased ratio in the valence quark region ($\Xbj \geq 0.3$) where the intermediate vector boson is predominantly interacting with $d$ quarks. The ratios corrected for non-isoscalar effects are in Figure \ref{fig:crossSectionsIso}. 

There is a weak preference for a smaller than predicted Pb/CH ratio at low $\Xbj$. These data are consistent with a previously published \minerva inclusive analysis \cite{nukePRL}: a deficit relative to the simulation at low $\Xbj$ that increases as the size of the nucleus increases. The mean $\Xbj$ and $Q^{2}$ of data events in the lowest bin are approximately $0.07$ and $2.0$ \GevSq, respectively. The amount of shadowing observed at this $\Xbj$ and $Q^{2}$ contrasts with charged-lepton scattering fits, which predict a ratio of 1.03 for Pb to CH \cite{Bodek:2010km}.

The ratios of carbon, iron, and lead to scintillator agree well with the simulation in the largest $\Xbj$ bin, $0.4 \leq \Xbj < 0.75$. This bin corresponds to the region where the EMC effect is dominant. The current resolution of the data is not sufficient to measure the EMC effect between the different nuclei at the level observed in charged-lepton data \cite{slacRatios}. The data likewise imply the differences between the EMC effect in charged leptons and neutrinos must be smaller than the  current \minerva data can resolve.

The ratios of total DIS cross sections as a function of $E_{\nu}$ for C, Fe and Pb to CH are shown in Fig.~\ref{fig:crossSections} (right). The ratio corrected for non-isoscalar effects is included in Figure \ref{fig:crossSectionsIso}. A smaller-than-expected ratio in the higher-energy bins of the lead to CH cross-section ratio is observed. This is consistent with the deficit in the lower $\Xbj$ bins, as the higher energy neutrino events will tend to have a higher $E_{had}$ and a lower $\Xbj$. In contrast, the ratio of C to CH at low energy is larger than unity with a large uncertainty consistent with the MC ratio of about 1.1. This is observed in the $\Xbj$ ratios as well, where the data ratio is larger than the simulated ratios in all bins.

Isoscalar corrections are applied to the data and simulation to correct for the difference in the per-nucleon cross section of two nuclei due to the difference in the way the neutrino interacts with the bound protons and neutrons. The isoscalar correction factors out this neutron excess. \textsc{genie} is used to predict the free-nucleon cross sections. As \minerva measures the ratio of cross section of different nuclei (C, Fe, Pb) to that of CH, the isoscalar correction becomes:

\begin{eqnarray}
	f_{\text{iso}} =  \left(\frac{A}{13}\right) \frac{7 \sigma (p_f)  + 6 \sigma (n_f)}{ Z_{A} \sigma (p_f) + N_{A} \sigma (n_f)},
	\label{eq:IsoCorNS}
\end{eqnarray}

\noindent where $A$ is the atomic number, $Z_{A}$ is the number of protons, $N_{A}$ is the number of neutrons, $\sigma (p_f)$ is the free proton cross section, and $\sigma (n_f)$ is the free neutron cross section. 

This correction does not take $\Xbj$-dependent partonic effects into account, and assumes the bound nuclear cross section is the same for all $A$. Isoscalar-corrected ratios as a function of $E_{\nu}$ and $\Xbj$ are shown in Fig. \ref{fig:crossSectionsIso}. Differences between the simulation and unity in the ratios stem from under-predicted CH backgrounds which are covered by the added uncertainty.

\begin{figure}
\centering
\ifnum\DoPrePrint=0
\includegraphics[trim = 0mm 10mm 7mm 8mm, clip, width=0.75\columnwidth]{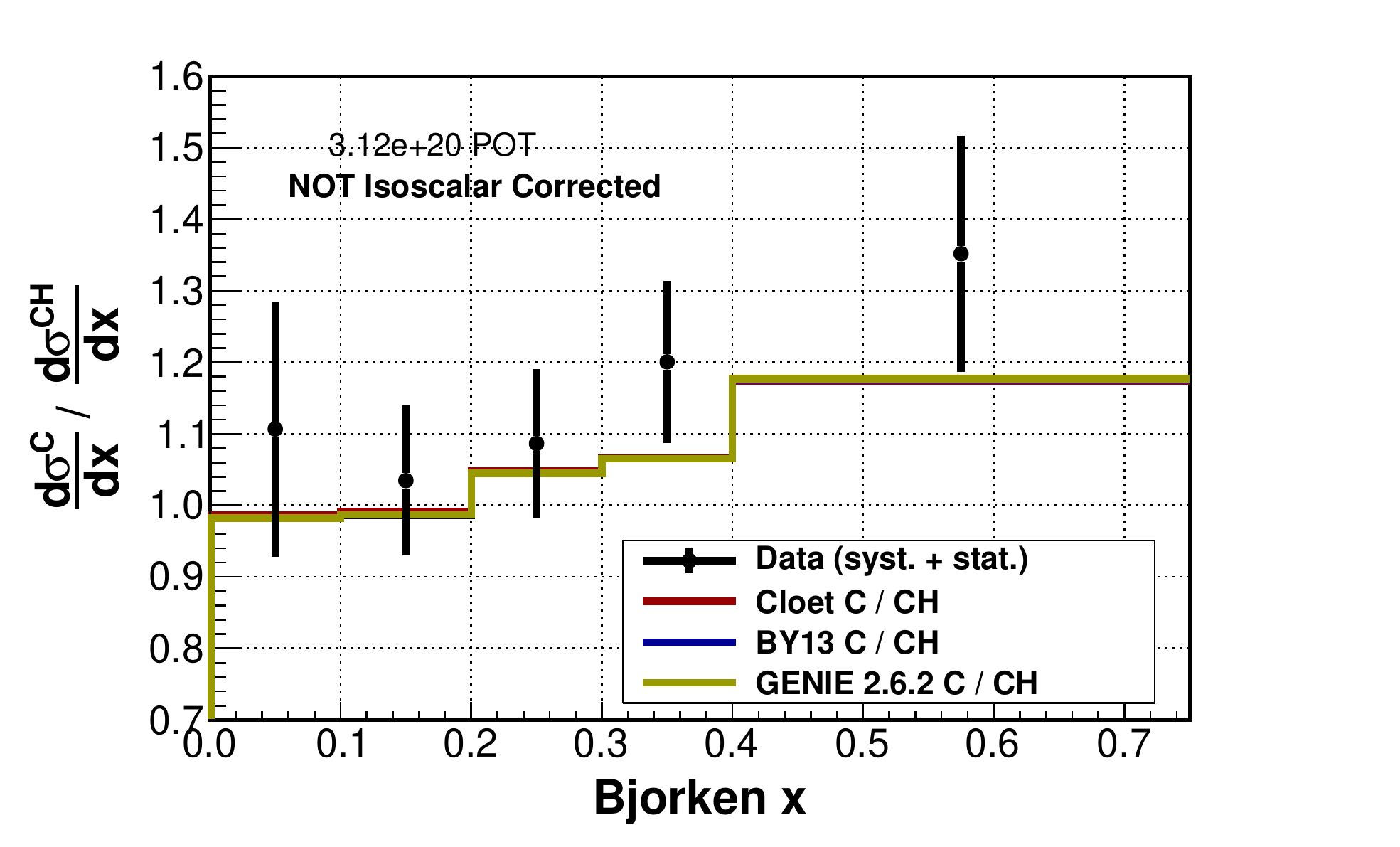}
\includegraphics[trim = 0mm 10mm 7mm 8mm, clip, width=0.76\columnwidth]{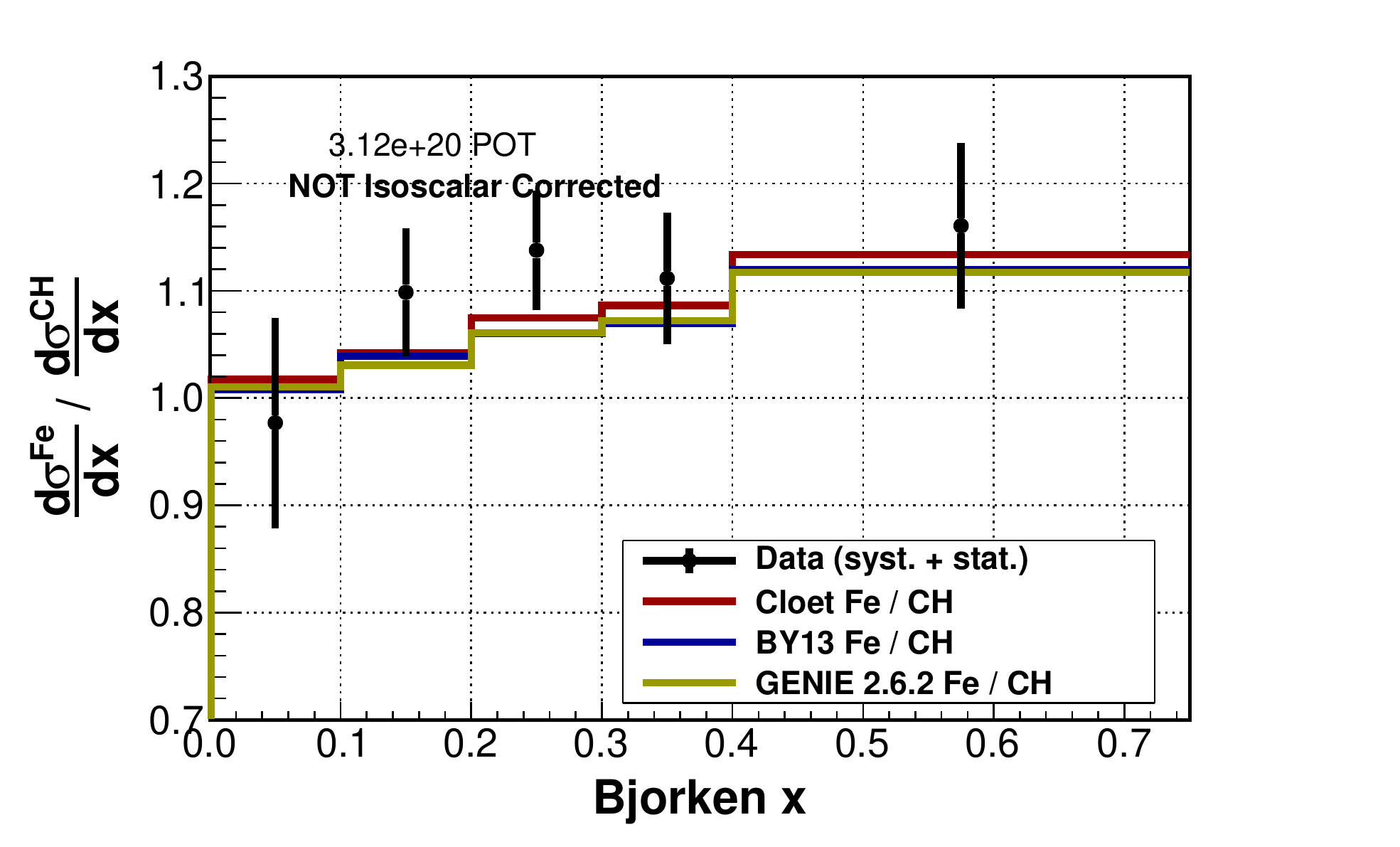}
\includegraphics[trim = 0mm 0mm 7mm 8mm, clip, width=0.77\columnwidth]{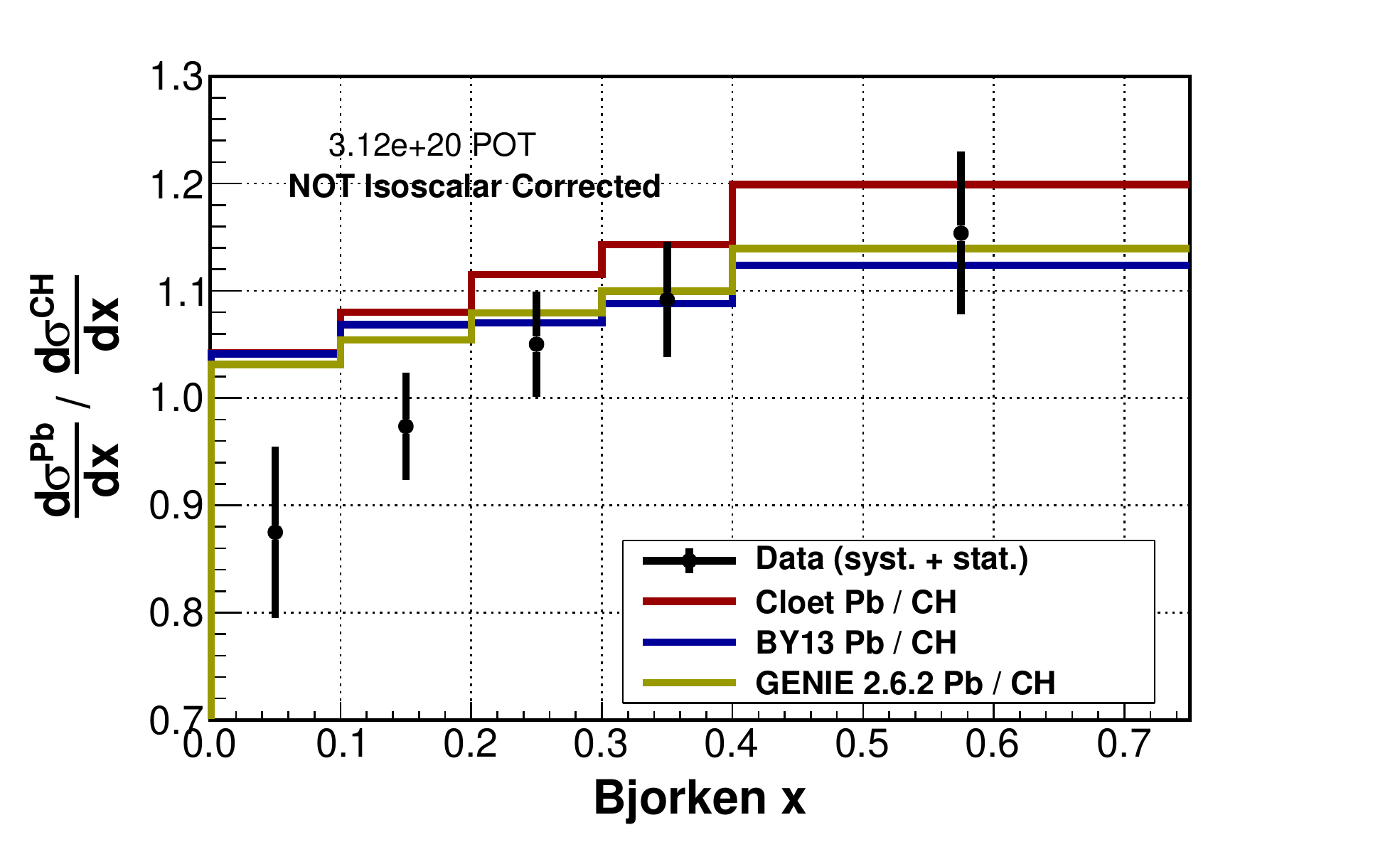}
\else
\includegraphics[trim = 0mm 10mm 7mm 8mm, clip, width=0.70\columnwidth]{MultiModel_CrossSectionRatio_x_z06_91_Nu_v22_prl.pdf}
\includegraphics[trim = 0mm 10mm 7mm 8mm, clip, width=0.71\columnwidth]{MultiModel_CrossSectionRatio_x_z26_92_Nu_v22_prl.pdf}
\includegraphics[trim = 0mm 0mm 7mm 8mm, clip, width=0.72\columnwidth]{MultiModel_CrossSectionRatio_x_z82_93_Nu_v22_prl.pdf}
\fi
\caption{DIS cross section ratios as a function of \Xbj for \minerva data (points) and various parameterizations of $\Xbj$-dependent nuclear effects \cite{genie, Bodek:2010km, cloet}. Note that the Cloet valence-quark model predictions are only valid for $\Xbj \geq 0.3 $. The error bars on the data are the combined statistical and systematic uncertainties. \label{fig:modelCompare}}
\end{figure}

The non-isoscalar corrected data are compared with non-isoscalar corrected alternative parameterizations of partonic nuclear effects applied to \textsc{genie} in Fig. \ref{fig:modelCompare}. The 2013 version of Bodek-Yang (BY13) \cite{Bodek:2010km} updates the parton distribution functions (PDFs) used in Bodek-Yang 2003 to include an $A$-dependent parameterization of the $\Xbj$-dependent effects based on charged-lepton scattering data. This parameterization uses updated data from the experiments \cite{bydata1, bydata2, bydata3, bydata4}. The Cloet model consists of an independent calculation of $F_{2}$ and $xF_{3}$ based on a convolution of the Nambu-Jona-Lasinio \cite{njl} nuclear wave function with free-nucleon valence PDFs \cite{cloet}, and does not include shadowing and anti-shadowing effects that dominate the $\Xbj \leq 0.3 $ kinematic region. The ratio calculation for the Cloet prediction assumes the Callan-Gross relationship $2\Xbj F_{1} = F_{2}$. Both BY13 and Cloet models have been shown to agree with charged-lepton DIS data in the EMC region. 

While the data do not currently have the sensitivity to distinguish between the different models at higher $\Xbj$, the deficit in data observed in the smallest $\Xbj$ bin cannot be explained by the updated Bodek-Yang model, the only model which is applicable at low $\Xbj$. The disagreement may be explained by the fact that BY13 contains a fit based on charged-lepton scattering which only contains a vector current. For a given $\Xbj$ and $Q^{2}$, the coherence length of hadronic fluctuations may be longer for the axial-vector current than the vector current \cite{cohLength}. This would allow shadowing to occur for neutrino scattering in the lowest $\Xbj$ bin where vector-current shadowing is greatly suppressed.

\begingroup
\squeezetable
\begin{table}
\begin{tabular}{ccccccccc}
\hline \hline
\multicolumn{9}{c}{Carbon}\\
\hline
$\Xbj$ & I & II & III & IV & V & VI & VII & Total \\ 
\hline
0.00--0.10 & 13.6 & 2.6 & 6.8 & 3.9 & 4.5 & 4.0 & 3.3 & 17.4 \\ 
0.10--0.20 & 7.3 & 4.2 & 3.6 & 1.3 & 3.8 & 1.6 & 1.8 & 10.3 \\ 
0.20--0.30 & 6.9 & 3.9 & 3.9 & 2.1 & 3.5 & 2.8 & 1.4 & 10.2 \\ 
0.30--0.40 & 8.0 & 0.6 & 5.4 & 3.5 & 3.3 & 1.4 & 1.4 & 11.0 \\ 
0.40--0.75 & 11.5 & 5.6 & 8.0 & 3.1 & 3.5 & 1.2 & 1.6 & 15.9 \\ 
\hline \hline
\multicolumn{9}{c}{Iron}\\
\hline
$\Xbj$ & I & II & III & IV & V & VI & VII & Total \\ 
\hline
0.00--0.10 & 6.3 & 1.7 & 3.6 & 3.4 & 3.3 & 4.1 & 1.9 & 10.0 \\ 
0.10--0.20 & 3.6 & 1.2 & 1.9 & 1.4 & 2.9 & 1.4 & 1.7 & 5.8 \\ 
0.20--0.30 & 3.4 & 0.1 & 1.9 & 1.1 & 2.8 & 1.1 & 1.8 & 5.4 \\ 
0.30--0.40 & 3.7 & 1.0 & 2.6 & 1.6 & 2.8 & 1.2 & 1.9 & 6.0 \\ 
0.40--0.75 & 5.0 & 1.9 & 3.6 & 2.3 & 2.7 & 0.7 & 1.8 & 7.7 \\ 
\hline \hline
\multicolumn{9}{c}{Lead}\\
\hline
$\Xbj$ & I & II & III & IV & V & VI & VII & Total \\ 
\hline
0.00--0.10 & 5.8 & 1.5 & 3.5 & 2.5 & 2.5 & 2.0 & 2.5 & 8.4 \\ 
0.10--0.20 & 3.2 & 1.1 & 1.8 & 0.8 & 2.4 & 1.6 & 1.8 & 5.2 \\ 
0.20--0.30 & 3.1 & 0.2 & 1.8 & 0.9 & 2.6 & 1.2 & 1.7 & 5.0 \\ 
0.30--0.40 & 3.4 & 0.3 & 2.4 & 1.3 & 2.5 & 0.9 & 1.5 & 5.4 \\ 
0.40--0.75 & 4.8 & 1.5 & 3.4 & 1.9 & 3.3 & 1.8 & 1.5 & 7.6 \\ 
\hline \hline
\end{tabular}
\caption{Uncertainties as a percentage on the ratio of DIS differential cross sections $\frac{d\sigma^{A}}{d \Xbj}/\frac{d\sigma^{CH}}{d \Xbj}$ for carbon (top), iron (center) and lead (bottom) with respect to $\Xbj$. The uncertainties are grouped by (I) data statistics, (II) CH background subtraction, (III) MC statistics, (IV) detector response to muons and hadrons (V) neutrino interactions, (VI) final-state interactions, and (VII) flux and target number. The rightmost column shows the total uncertainty due to all sources.}
\label{tab:x_ratio_sys_errors}
\end{table}
\endgroup

Neutrino-nucleus DIS presents a novel method to measure partonic nuclear effects in the weak sector. \minerva has measured this process using a variety of nuclear targets for the first direct measurement of neutrino-nuclear effects by isolating a region of high-$Q^{2}$ and high-$W$ events ($Q^{2} \geq 1.0$ \GevSq and $W \geq 2.0$ \Gev). The measured cross-section ratios show a general trend of being larger than the simulation for the lightest nucleus (C). Conversely, the data fall below the simulation in the heaviest nucleus (Pb) at high energy and low $\Xbj$, a trend also observed in a previous Letter \cite{nukePRL}. The data agree with \textsc{genie}'s treatment of the EMC effect between $\Xbj = 0.3$ and $\Xbj = 0.75$. The lower than expected Pb / CH ratio at large neutrino energy ($E_{\nu} > 20$ GeV) and low Bjorken-x ($\Xbj < 0.1$) is consistent with calculations \cite{Kopeliovich:2012kw} predicting a different kinematic threshold for shadowing in neutrino nucleus compared to charged-lepton nucleus scattering. Future studies with \minerva will utilize a higher-energy neutrino spectrum, and will be able to probe this interesting shadowing region by reducing the average \Xbj of neutrino DIS events. 

\ifnum\sizecheck=0
\input acknowledgement.tex   % input acknowledgement
\input{biblio.tex} % input references
\fi

%\ifnum\PRLsupp=1
%\clearpage
%\onecolumngrid
%\input{appendix.tex}
%\fi

\end{document}

%% file: authorlist.tex
\newcommand{\Rutgers}{Rutgers, The State University of New Jersey, Piscataway, New Jersey 08854, USA}
\newcommand{\Hampton}{Hampton University, Dept. of Physics, Hampton, VA 23668, USA}
\newcommand{\Dortmund}{Institute of Physics, Dortmund University, 44221, Germany }
\newcommand{\Otterbein}{Department of Physics, Otterbein University, 1 South Grove Street, Westerville, OH, 43081 USA}
\newcommand{\JMU}{James Madison University, Harrisonburg, Virginia 22807, USA}
\newcommand{\Florida}{University of Florida, Department of Physics, Gainesville, FL 32611}
\newcommand{\UCIrvine}{Department of Physics and Astronomy, University of California, Irvine, Irvine, California 92697-4575, USA}
\newcommand{\CBPF}{Centro Brasileiro de Pesquisas F\'{i}sicas, Rua Dr. Xavier Sigaud 150, Urca, Rio de Janeiro, Rio de Janeiro, 22290-180, Brazil}
\newcommand{\PUCP}{Secci\'{o}n F\'{i}sica, Departamento de Ciencias, Pontificia Universidad Cat\'{o}lica del Per\'{u}, Apartado 1761, Lima, Per\'{u}}
\newcommand{\INRM}{Institute for Nuclear Research of the Russian Academy of Sciences, 117312 Moscow, Russia}
\newcommand{\Jlab}{Jefferson Lab, 12000 Jefferson Avenue, Newport News, VA 23606, USA}
\newcommand{\Pittsburgh}{Department of Physics and Astronomy, University of Pittsburgh, Pittsburgh, Pennsylvania 15260, USA}
\newcommand{\Guanajuato}{Campus Le\'{o}n y Campus Guanajuato, Universidad de Guanajuato, Lascurain de Retana No. 5, Colonia Centro, Guanajuato 36000, Guanajuato M\'{e}xico.}
\newcommand{\Athens}{Department of Physics, University of Athens, GR-15771 Athens, Greece}
\newcommand{\Tufts}{Physics Department, Tufts University, Medford, Massachusetts 02155, USA}
\newcommand{\WM}{Department of Physics, College of William \& Mary, Williamsburg, Virginia 23187, USA}
\newcommand{\FNAL}{Fermi National Accelerator Laboratory, Batavia, Illinois 60510, USA}
\newcommand{\Purdue}{Department of Chemistry and Physics, Purdue University Calumet, Hammond, Indiana 46323, USA}
\newcommand{\MCLA}{Massachusetts College of Liberal Arts, 375 Church Street, North Adams, MA 01247}
\newcommand{\UMD}{Department of Physics, University of Minnesota -- Duluth, Duluth, Minnesota 55812, USA}
\newcommand{\Northwestern}{Northwestern University, Evanston, Illinois 60208}
\newcommand{\UNI}{Universidad Nacional de Ingenier\'{i}a, Apartado 31139, Lima, Per\'{u}}
\newcommand{\Rochester}{University of Rochester, Rochester, New York 14627 USA}
\newcommand{\Austin}{Department of Physics, University of Texas, 1 University Station, Austin, Texas 78712, USA}
\newcommand{\USM}{Departamento de F\'{i}sica, Universidad T\'{e}cnica Federico Santa Mar\'{i}a, Avenida Espa\~{n}a 1680 Casilla 110-V, Valpara\'{i}so, Chile}
\newcommand{\Geneva}{University of Geneva, 1211 Geneva 4, Switzerland}
\newcommand{\Chicago}{Enrico Fermi Institute, University of Chicago, Chicago, IL 60637 USA}
\newcommand{\hired}{}
\newcommand{\OregonState}{Department of Physics, Oregon State University, Corvallis, Oregon 97331, USA}
\newcommand{\bmeThanks}{now at SLAC National Accelerator Laboratory, Stanford, California 94309 USA}
\newcommand{\higueraThanks}{University of Houston, Houston, Texas, 77204, USA}
\newcommand{\damartinezThanks}{Now at Illinois Institute of Technology}
\newcommand{\joelmousseauThanks}{now at University of Michigan, Ann Arbor, MI, 48109}
\newcommand{\LazaThanks}{also at Department of Physics, University of Antananarivo, Madagascar}
\newcommand{\twaltonThanks}{now at Fermi National Accelerator Laboratory, Batavia, IL USA 60510}
\newcommand{\jwolcottThanks}{Now at Tufts University, Medford, Massachusetts 02155, USA }

% 74 total signatories.
\author{J.~Mousseau}\thanks{\joelmousseauThanks}  \affiliation{\Florida}
\author{M.Wospakrik}                      \affiliation{\Florida}
\author{L.~Aliaga}                        \affiliation{\WM}
\author{O.~Altinok}                       \affiliation{\Tufts}
\author{L.~Bellantoni}                    \affiliation{\FNAL}
\author{A.~Bercellie}                     \affiliation{\Rochester}
\author{M.~Betancourt}                    \affiliation{\FNAL}
\author{A.~Bodek}                         \affiliation{\Rochester}
\author{A.~Bravar}                        \affiliation{\Geneva}
\author{H.~Budd}                          \affiliation{\Rochester}
\author{T.~Cai}                           \affiliation{\Rochester}
\author{M.F.~Carneiro}                    \affiliation{\CBPF}
\author{M.E.~Christy}                     \affiliation{\Hampton}
\author{J.~Chvojka}                       \affiliation{\Rochester}
\author{H.~da~Motta}                      \affiliation{\CBPF}
\author{J.~Devan}                         \affiliation{\WM}
\author{S.A.~Dytman}                      \affiliation{\Pittsburgh}
\author{G.A.~D\'{i}az~}                   \affiliation{\Rochester}  \affiliation{\PUCP}
\author{B.~Eberly}\thanks{\bmeThanks}     \affiliation{\Pittsburgh}
\author{J.~Felix}                         \affiliation{\Guanajuato}
\author{L.~Fields}                        \affiliation{\FNAL}  \affiliation{\Northwestern}
\author{R.~Fine}                          \affiliation{\Rochester}
\author{A.M.~Gago}                        \affiliation{\PUCP}
\author{R.Galindo}                        \affiliation{\USM}
\author{H.~Gallagher}                     \affiliation{\Tufts}
\author{A.~Ghosh}                         \affiliation{\CBPF}  \affiliation{\CBPF}
\author{T.~Golan}                         \affiliation{\Rochester}  \affiliation{\FNAL}
\author{R.~Gran}                          \affiliation{\UMD}
\author{D.A.~Harris}                      \affiliation{\FNAL}
\author{A.~Higuera}\thanks{\higueraThanks}  \affiliation{\Rochester}  \affiliation{\Guanajuato}
\author{K.~Hurtado}                       \affiliation{\CBPF}  \affiliation{\UNI}
\author{M.~Kiveni}                        \affiliation{\FNAL}
\author{J.~Kleykamp}                      \affiliation{\Rochester}
\author{M.~Kordosky}                      \affiliation{\WM}
\author{T.~Le}                            \affiliation{\Tufts}  \affiliation{\Rutgers}
\author{E.~Maher}                         \affiliation{\MCLA}
\author{S.~Manly}                         \affiliation{\Rochester}
\author{W.A.~Mann}                        \affiliation{\Tufts}
\author{C.M.~Marshall}                    \affiliation{\Rochester}
\author{D.A.~Martinez~Caicedo}\thanks{\damartinezThanks}  \affiliation{\FNAL}
\author{K.S.~McFarland}                   \affiliation{\Rochester}  \affiliation{\FNAL}
\author{C.L.~McGivern}                    \affiliation{\Pittsburgh}
\author{A.M.~McGowan}                     \affiliation{\Rochester}
\author{B.~Messerly}                      \affiliation{\Pittsburgh}
\author{J.~Miller}                        \affiliation{\USM}
\author{A.~Mislivec}                      \affiliation{\Rochester}
\author{J.G.~Morf\'{i}n}                  \affiliation{\FNAL}
\author{D.~Naples}                        \affiliation{\Pittsburgh}
\author{J.K.~Nelson}                      \affiliation{\WM}
\author{A.~Norrick}                       \affiliation{\WM}
\author{Nuruzzaman}                       \affiliation{\Rutgers}  \affiliation{\USM}
\author{J.~Osta}                          \affiliation{\FNAL}
\author{V.~Paolone}                       \affiliation{\Pittsburgh}
\author{J.~Park}                          \affiliation{\Rochester}
\author{C.E.~Patrick}                     \affiliation{\Northwestern}
\author{G.N.~Perdue}                      \affiliation{\FNAL}  \affiliation{\Rochester}
\author{L.~Rakotondravohitra}\thanks{\LazaThanks}  \affiliation{\FNAL}
\author{M.A.~Ramirez}                     \affiliation{\Guanajuato}
\author{R.D.~Ransome}                     \affiliation{\Rutgers}
\author{H.~Ray}                           \affiliation{\Florida}
\author{L.~Ren}                           \affiliation{\Pittsburgh}
\author{D.~Rimal}                         \affiliation{\Florida}
\author{P.A.~Rodrigues}                   \affiliation{\Rochester}
\author{D.~Ruterbories}                   \affiliation{\Rochester}
\author{H.~Schellman}                     \affiliation{\OregonState}  \affiliation{\Northwestern}
\author{D.W.~Schmitz}                     \affiliation{\Chicago}  \affiliation{\FNAL}
\author{C.J.~Solano~Salinas}              \affiliation{\UNI}
\author{N.~Tagg}                          \affiliation{\Otterbein}
\author{B.G.~Tice}                        \affiliation{\Rutgers}
\author{E.~Valencia}                      \affiliation{\Guanajuato}
\author{T.~Walton}\thanks{\twaltonThanks}  \affiliation{\Hampton}
\author{J.~Wolcott}\thanks{\jwolcottThanks}  \affiliation{\Rochester}

\author{G.~Zavala}                        \affiliation{\Guanajuato}
\author{D.~Zhang}                         \affiliation{\WM}

\collaboration{The MINER$\nu$A Collaboration}\ \noaffiliation
\date{\today}

%% file: acknowledgement.tex
This work was supported by the Fermi National Accelerator Laboratory under US Department of Energy contract No. DE-AC02-07CH11359 which included the MINERvA construction project. Construction support was also granted by the United States National Science Foundation under Award PHY-0619727 and by the University of Rochester. Support for participating scientists was provided by NSF and DOE (USA), by CAPES and CNPq (Brazil), by CoNaCyT (Mexico), by CONICYT (Chile), by CONCYTEC, DGI-PUCP and IDI/IGI-UNI (Peru), by Latin American Center for Physics (CLAF). We thank the MINOS Collaboration for use of its near detector data. We acknowledge the dedicated work of the Fermilab staff responsible for the operation and maintenance of the beamline, detector, and the computing infrastructure.